\documentclass{jfm}
\usepackage{amsmath}
\usepackage{graphicx}
\usepackage{color}
\usepackage{cancel}

\newcommand{\lra}[1]{\langle #1 \rangle }

\newcommand{\bds}[1]{\boldsymbol{#1}}
\newcommand{\bfnabla}{\boldsymbol{\nabla}}

\shorttitle{A Lagrangian pdf model for collisional turbulent fluid--particle flows}
\shortauthor{A.\ Innocenti, R.\ O.\ Fox, M.\ V.\ Salvetti and S.\ Chibbaro,}

\title{A Lagrangian probability-density-function model for collisional turbulent fluid--particle flows. II. Application to homogeneous flows}

\author{A.~Innocenti\aff{1}\aff{2},
   R.~O.~Fox\aff{3}
, M.~V.~Salvetti\aff{2}  \and S.~Chibbaro{\aff{1}
  \corresp{\email{sergio.chibbaro@upmc.fr}}}}

\affiliation{\aff{1}Sorbonne Universit\'e, UPMC Univ Paris 06, CNRS, UMR 7190, Institut Jean Le Rond d'Alembert, F-75005 Paris, France
\aff{2}Dipartimento di Ingegneria Civile e Industriale, Universit\`a di Pisa, Via G. Caruso 8, 56122 Pisa, Italia
\aff{3}Department of Chemical and Biological Engineering, 618 Bissell Road, Iowa State University, Ames, IA 50011-1098, USA}

\begin{document}
\maketitle
\begin{abstract}
The Lagrangian probability-density-function model, proposed in Part I for dense particle-laden turbulent flows, is validated here against Eulerian-Lagrangian direct numerical simulation (EL) data for different homogeneous flows, namely statistically steady and decaying homogeneous isotropic turbulence, homogeneous-shear flow and cluster-induced turbulence (CIT).
We consider the general model developed in Part I adapted to the homogeneous case together with a simplified version in which the decomposition of the phase-averaged (PA) particle-phase  fluctuating energy into the spatially correlated and uncorrelated components
is not used, and only total exchange of kinetic energy between phases is allowed. The simplified  model employs the standard two-way coupling approach.
The comparison between EL simulations and the two stochastic models in homogeneous and isotropic turbulence and in homogeneous-shear flow shows that in all cases both models are capable to reproduce rather well the flow behaviour, notably for dilute flows.
The analysis of the CIT gives more insights on the physical nature of such systems and about the quality of the models.
Results elucidate the fact that simple two-way coupling is sufficient to induce turbulence, even though the granular energy is not considered. Furthermore, first-order moments including velocity of the fluid seen by particles can be fairly well represented with such a simplified stochastic model. However, the decomposition into spatially correlated and uncorrelated components is found to be necessary to account for anisotropic energy exchanges.
When these factors are properly accounted for as in the complete model, the agreement with the EL statistics is satisfactory up to second order. 
\end{abstract}


\section{Introduction}
Particle-laden flows represent an important class of natural and industrial flows \citep{crowe2011multiphase}.
In many applications, these flows are heavily charged in particles \citep{stickel2005fluid,forterre2008flows,guazzelli2011physical} and are often turbulent \citep{balachandar2010turbulent}.
Given the complexity of such phenomena, to put forward a reduced model is mandatory for practical purposes.
To guide this development it is very useful to disentangle the different physical mechanisms at play, and in particular to understand how to cope with the effect of increasing the particle mass loading and the consequent growing importance of collisions and two-way coupling.
Unfortunately, it is hard to find clear-cut frontiers between the different regimes \citep{elghobashi1992direct},
and thus some heuristic considerations are always needed.

Generally speaking, two classes of modelling approaches can be chosen for turbulent flows, the Eulerian and the Lagrangian ones \citep{Pope_turbulent}.
When the flow is dilute or moderately dense, the Lagrangian approach is mature \citep{minier2014guidelines} and has been found to be superior in many cases \citep{Pei_06}. 
On the other hand, when the volume occupied by the particles is relatively large, collisions are completely dominant, the matter becomes granular, turbulence is absent and a hydrodynamic approach is often natural \citep{puglisi2014transport}. 
We consider in this work the intermediate regime, which is less clear.
From a historical perspective, turbulence models for dense flows have been developed in an Eulerian framework on a purely heuristic grounds in analogy with single-phase models \citep{elghobashi1983two,viollet1994modelling}.
Only recently, it has been shown that most of those models suffer from some drawbacks and a more rigorous approach has been followed to 
formulate a complete Reynolds-stress model \citep{Fox2014}. 
In Part~I, we have developed a Lagrangian pdf approach, which leads to a stochastic model for the particle phase, and which has to be coupled with a consistent Reynolds-stress model for the fluid.
This approach permits to solve directly for the velocity of the fluid seen by particles, which has instead to be modelled in the Eulerian approach. 
This can be an advantage for modelling purposes and mainly its knowledge can be important in some applications.

From a modelling point of view, a key but often overlooked point for the modelling of turbulent dense flows is the separation between the spatially correlated part, contributing to the turbulent kinetic energy, and the uncorrelated part, responsible for the granular temperature \citep{dasgupta1994turbulent,fevrier2005partitioning}. They sum up, of course, to the total energy $\kappa_p = k_p + \frac{3}{2} \langle \Theta_p \rangle$ where $\langle \Theta_p \rangle$ is the granular temperature. If collisions are absent, the need to decompose the particle velocity into spatially correlated and uncorrelated components is less obvious. However, owing to the fact that particle--particle collisions are driven by the spatially uncorrelated velocity component, this decomposition is thought to be crucial for collisional flows.

In this work, we test two stochastic Lagrangian models describing the particle phase, coupled with Reynolds-average Eulerian equations for the fluid phase.
The first model, derived in Part~I, is based on velocity partitioning between correlated and uncorrelated components. The second one is a simplified version, where only the total particle velocity, derived as the sum of the two component, is resolved, leading to the lack of distinction between the particle turbulent kinetic energy and the granular temperature. 
We focus here on statistically homogeneous turbulence.
In particular, the goal of the paper is to understand if the stochastic models are able to deal with the momentum and energy exchange between phases, and the particle concentration fluctuations. 
These ingredients are essential in all moderately dense particle-laden flows, and therefore it is important to 
use the homogeneous cases in order to isolate 
their modelling from other complex features present in non-homogeneous configurations (e.g.\ spatial fluxes).
In particular, we are interested in cluster-induced turbulence (CIT), which occurs in fluid--particle flows when (i) the mean mass loading $\varphi$, defined
by the ratio of the specific masses of the particle and fluid phases, is of order
one or larger; and (ii) the difference between the mean phase velocities is non-zero.
Interestingly, in statistically stationary flows, fluctuations in particle concentration can generate
and sustain fluid-phase turbulence, which we refer to as fully developed CIT. 
Given that the density ratio $\frac{\rho_p}{\rho_f}$ is very large in gas--particle flows, CIT is
ubiquitous in practical engineering and environmental flows when body forces or
inlet conditions generate a mean velocity difference, such as the gravity-driven flows
studied herein. 
Some fundamental properties of such flows 
has been recently studied via Eulerian--Lagrangian numerical simulations \citep{capecelatro2014numerical,Fox2015},
which will be used for comparison.
Notably, the use of models can be relevant to emphasise the mechanisms underlying the volume-fraction fluctuations.

From a physical point of view, 
we want to assess the stochastic models with respect to their ability to reproduce the statistical features of both the particle and fluid phases at high mass loading.
It is well known that turbulent particle-laden flows
 in the dilute limit, where the fluid-phase turbulence interacts with inertial particles without
significant feedback from the particles,
 display a preferential concentration of particles in certain regions \citep{balkovsky2001intermittent,balachandar2010turbulent}.
 In particular, 
it is well established that dilute suspensions of heavy particles
in isotropic turbulence will preferentially concentrate in regions of high strain rate
and low vorticity \citep{eaton1994preferential}.
When two-way coupling between the phases is non-negligible, additional
effects may be responsible for enhancing the settling rate and spatial segregation of
the particles. 
Among the possible effects,
it is worth mentioning the enhancement 
of particle settling velocity with increasing volume fraction \citep{bosse2006small}, and the creation of strong anisotropy due to the crossing trajectory effect \citep{ferrante2003physical} causing also a drag reduction.
However, even more impressive is the situation at higher mass-loading, where the fluctuating segregation of particles,  together with collisions, have been found to create large cluster that induce turbulence in a fluid otherwise at rest \citep{glasser1998bubbles,Fox2015}, giving rise to CIT.
Notably, in gravity-driven CIT, particles accumulate in regions of low vorticity, as is seen in classical preferential concentration of low-mass-loading
suspensions. However, in CIT the vorticity is generated in shear layers between
clusters, unlike in classical preferential concentration, where vorticity would exist
even in the absence of the disperse phase. 

The goal of the present work is precisely to understand whether the stochastic model proposed in Part~I is capable of reproducing quantitatively the phenomena observed in homogeneous particle-laden flows, and also to find out which elements are necessary to trigger the instability leading to CIT.
The paper is organised as follows. In \S\ref{sec::Lpdf} we briefly review the key features of the Lagrangian pdf models developed in Part~I for the case of statistically homogeneous flows. Then, in \S\ref{sec::NR}, the models are applied to increasingly more complex particle-laden flows and the results compared to data from the literature. In \S\ref{sec::Con}, conclusions are drawn concerning the relative merits of the proposed models, along with a discussion of future challenges to be faced with applying them to spatially inhomogeneous flows.

\section{Lagrangian pdf model for particle-laden flows}\label{sec::Lpdf}

In Part~I of this work, we have developed the general formalism for the Lagrangian pdf approach to dense flows, and we have proposed a rather general stochastic model, which should be suitable for moderately dense flows where collisions play a role but are not completely dominant.  
We present in \S\ref{sec:part-model} the model for the case of homogeneous flows dealt with in this Part~II. For the particle phase, we propose also a second simplified model, which takes into account the exchanges between the phases, but not the collisions and does not distinguish between the correlated and uncorrelated parts of the particle-phase velocity field.

\subsection{Stochastic model for particle phase}
\label{sec:part-model}

The set of stochastic equations for the particle phase, expressed for a homogeneous flow, is detailed in \eqref{eq:SDEp-homo-coll-first}--\eqref{eq:SDEp-homo-coll-fourth} below.
\begin{equation} \label{eq:SDEp-homo-coll-first}
d  { x}_{p,i}  = { V}_{p,i} \, dt = ( U_{p,i} + \delta v_{p,i}) \, dt 
\end{equation}
where $ {\boldsymbol{ x}}_p$ is the particle position and ${\bf V}_p$ is the particle velocity. As explained in Part~I, following \cite{fevrier2005partitioning} and \cite{Fox2015}, the particle velocity is decomposed in a spatially correlated part ${\bf U}_p$, and in a uncorrelated residual, $\delta {\bf v}_p$. The former is governed by
\begin{equation} \label{eq:SDEp-homo-coll-second}
\begin{array}{lcl}
d  { U}_{p,i} & = & \displaystyle{\frac{{ U}_{s,i} - { U}_{p,i}}{\tau_p} }\, dt + { g}_i \, dt - \displaystyle{\frac{1}{\langle \alpha_p \rangle \rho_p} \frac{\partial \langle \alpha_p \rangle \rho_p \langle P_{ij} \rangle}{\partial x_j}} + {\delta} { v}_{p,j} \displaystyle{\frac{\partial \langle { U}_{p,i} \rangle}{\partial x_j}} \, dt  \\\\ & & - \frac{1}{T_{Lp}} ({ U}_{p,i} - \langle { U}_{p,i} \rangle ) \, dt + \sqrt{{C_{p}} \varepsilon_p} \; d{ W}_{p,i} .
\end{array}
\end{equation}
The first term of the RHS of \eqref{eq:SDEp-homo-coll-second} is the drag force related to the correlated part of the particle velocity, in which ${\bf U}_s$ is the fluid velocity seen by the particle and $\tau_p$ the particle relaxation time (hereinafter taken as a constant). The second term is the effect of gravity, $\bf g$, while the third is a pressure term, in which $\rho_p$ is the particle density, $\alpha_p$ the particle-phase volume fraction and $\langle P_{ij} \rangle = \langle \delta {v}_{p,i} \delta {v}_{p,j} \rangle$ is the particle-phase pressure tensor. The brackets $\langle \cdot \rangle$ denote phase-specific Reynolds average. The fourth and fifth terms are production and relaxation, respectively, in which $T_{Lp}$ is the particle Lagrangian time scale (defined in the following). Finally, the last contribution is a diffusion term, in which $C_{p}$ is a model constant to be a priori assigned, $\varepsilon_p$ is the particle dissipation and $d{ W}_{p,i}$ is a Wiener stochastic process.

The uncorrelated residual velocity is modelled by
\begin{equation} \label{eq:SDEp-homo-coll-third}
\begin{array}{lcl}
d \, {\delta} { v}_{p,i} & = & - \displaystyle{\frac{{\delta} { v}_{p,i}}{\tau_p}} \, dt + \displaystyle{\frac{1}{\langle \alpha_p \rangle \rho_p} \frac{\partial \langle \alpha_p \rangle \rho_p \langle P_{ij} \rangle}{\partial x_j}} - {\delta} { v}_{p,j} \displaystyle{\frac{\partial \langle { U}_{p,i} \rangle}{\partial x_j}} \, dt  + B_{\delta,ij} \; d{ W}_{\delta,j}   \\\\ && - \displaystyle{\frac{(1+e)(3-e)}{4 \tau_c }} \delta v_{p,i} \, dt + \sqrt{\frac{1}{2\tau_c } (1+e)^2 \langle \Theta_p \rangle} \; dW_{c ,i} .
\end{array}
\end{equation}
The first four terms in the RHS of \eqref{eq:SDEp-homo-coll-third} are analogous to the ones in \eqref{eq:SDEp-homo-coll-second}. In particular, $d{ W}_{\delta}$ is a Wiener stochastic process and $B_{\delta}$ is a diffusion matrix, whose expression is given in the following. The last two terms take into account collisions; $e$ is a restitution coefficient, to be a priori specified, $ dW_{c}$ is another Wiener process and $\langle \Theta_p \rangle$ is the granular temperature, defined as 
$
 \langle \Theta_p \rangle = \frac{1}{3} \langle \delta {\bf v}_p \bcdot \delta {\bf v}_p \rangle .
$
In particular, the following relation holds for the fluctuating energy partitioning: $\kappa_p = k_p + \frac{3}{2} \langle \Theta_p \rangle$, 
where $\kappa_p = \frac{1}{2} \langle {\bf v}_p \bcdot {\bf v}_p \rangle$ is the total particle-phase fluctuating energy and $k_p = \frac{1}{2} \langle {\bf u}_p \bcdot {\bf u}_p \rangle$ the turbulent particle-phase kinetic energy, ${\bf v}_p$ and ${\bf u}_p$ being the fluctuations arising from the Reynolds decomposition of ${\bf V}_p$ and ${\bf U}_p$, respectively. 
Finally, $\tau_c$ is a characteristic time for collisions, having the following expression:
\begin{equation}
\tau_c = \frac{\sqrt{\pi} d_p }{6 C_c \langle \alpha_p \rangle \langle \Theta_p \rangle^{1/2} } ,
\end{equation}
$d_p$ being the particle diameter and $C_c$ a model parameter \citep{capecelatro2016}.

The model for the fluid seen by the particles is 
\begin{equation} \label{eq:SDEp-homo-coll-fourth}
\begin{array}{lcl}
d{ U}_{s,i}(t) & = & - \displaystyle{\frac{1}{\rho_f} \frac{\partial \langle p_f \rangle}{\partial x_i}} \, dt  + { G_{i,j} ({ U}_{s,j} - \langle {U}_{f,j} \rangle )} \,dt  - \varphi \left( \displaystyle{\frac{{ U}_{s,i} - { U}_{p,i}}{\tau_p}} \right) \, dt   + { g}_i \, dt \\ & & + \Bigl [ \varepsilon_f \Bigl ( C_{0f} {b}_i \frac{\tilde{k}_f}{k_f} + \frac{2}{3} \Bigl ( {b}_i \frac{\tilde{k}_f}{k_f} -1 \Bigl) \Bigl )  +  2 \varphi  \displaystyle{\frac{\langle { U}_{p,i} - { U}_{s,i} \rangle}{\tau_p}} (\langle { U}_{s,i} \rangle - \langle {U}_{f,i} \rangle) \\ & &\qquad  { - 2\frac{\langle \alpha_p \rangle}{\langle \alpha_f \rangle \rho_f}\frac{\partial \langle p_f \rangle}{\partial x_i} (\langle U_{s,i} \rangle - \langle U_{f,i}\rangle)} \Bigl ]^{{1}/{2}}   d{ W}_{s,i} .
\end{array}
\end{equation}
The first term of the RHS is the pressure gradient term, where $\rho_f$ is the fluid density and $p_f$ the fluid pressure; in general, the subscript $f$ denotes a flow variable in the fluid phase. The second term is a relaxation term, where
\begin{equation}
G_{ij} = -\frac{1}{T_{L,i}^*} \delta_{ij} + G_{ij}^a .
\end{equation} 
$T_{L,i}^*$ is a modified fluid time-scale, which takes into account the anisotropy of the flow and particle inertia, defined by
\begin{equation}\label{eq:T_L}
T_{L,i}^* = \frac{T_{Lf}}{\sqrt{1 + \zeta_i \beta^2 \frac{3 |\langle {\bf U}_r \rangle |^2}{2k_f}}}, \quad 
T_{Lf} = \frac{2}{\left( 1 + \frac{3}{2} C_{0f} \right)} \frac{k_f}{\varepsilon_f}
\end{equation}
where $\zeta_1 =1$ in the mean drift direction and $\zeta_{2,3} =4$ in the cross directions, $\beta = T_{Lf} / T_{Ef}$ is the ratio of the Lagrangian and the Eulerian timescales and ${\bf U}_r = {\bf U}_p - {\bf U}_s$ is the relative velocity.  $k_f$ and $\varepsilon_f$ are the fluid turbulent kinetic energy and dissipation. ${\bf G}^a$ is a traceless matrix to be added to generalize the model as shown in Part I:
\begin{equation}
G_{ij}^a = C_{2f} \frac{\partial \langle U_{f,i}\rangle}{\partial x_j} .
\end{equation}
It corresponds to the Isotropization-of-production contribution in the LRR-IP model, with $C_{2f}$ being the IP constant.
The value of the model constant $C_{0f}$ is established by the relation, see \citep{Pop_94a}:
\begin{equation}
C_{0f} = \frac{2}{3} \left( C_{Rf} - 1 +C_{2f} \frac{\mathcal{P}}{\varepsilon_f} \right) .
\end{equation}
where $C_{Rf}$ is the Rotta constant and $\mathcal{P}$ the mean shear production.
The third term in \eqref{eq:SDEp-homo-coll-fourth} accounts for two-way coupling, $\varphi$ being the mean mass loading, defined as
$
\varphi = \frac{\rho_p \langle \alpha_p \rangle}{\rho_f \langle \alpha_f \rangle} .
$
Finally, the last term is a stochastic diffusion process extended to dense flows in which ${b}_i = T_{Lf} / { T}_{L,i}^*$, 
\begin{equation}
\tilde{k}_f = \frac{3}{2} \frac{ \sum_{i=1}^3 { b}_i \langle ({ U}_{s,i} - \langle U_{f,i}\rangle )^2  \rangle}{\sum_{i=1}^3{ b}_i }
\end{equation} 
and $dW_s$ is an additional Wiener process.

When the correlation $\langle \delta v_{p,i}  \delta v_{p,j} \rangle$ is evaluated, the diffusion matrix $B_{\delta}$ must give the particle-phase Reynolds-stress tensor multiplied by the proper coefficient together with a diagonal isotropic part. Using a Choleski decomposition we obtain:
\begin{equation}
\begin{aligned}
& B_{\delta,11} = \left[ f_s \frac{\varepsilon_p}{k_p} \langle u_{p,1} u_{p,1} \rangle + (1-f_s) \frac{2}{3} \varepsilon_p \right]^{1/2},   \\
& B_{\delta,i1} = \frac{1}{B_{\delta,11}} f_s \frac{\varepsilon_p}{k_p} \langle u_{p,i} u_{p,1} \rangle , \quad  1 < i \le 3  \\
& B_{\delta,ii} = \left[ f_s \frac{\varepsilon_p}{k_p} \langle u_{p,i} u_{p,i} \rangle + (1-f_s) \frac{2}{3} \varepsilon_p - \sum_{j=1}^{i-1} B_{\delta,ij}^2 \right]^{1/2}, \quad 1 < i \le 3  \\
& B_{\delta,ij} = \frac{1}{B_{\delta,jj}} \left( f_s \frac{\varepsilon_p}{k_p} \langle u_{p,i} u_{p,j} \rangle - \sum_{k=1}^{j-1} B_{\delta,ik} B_{\delta,jk} \right), \quad 1 < j < i \le 3  \\
& B_{\delta,ij} = 0, \quad i < j \le 3 \, ;
\end{aligned}
\end{equation}
where $0 \le f_s \le 1$ is a parameter tuning the anisotropy of the particle dissipation tensor. The latter is defined as follows:
\begin{equation} \label{eq:distensor}
\bds{\varepsilon}_p =  \varepsilon_p \left[ f_s \frac{\langle {\bf u}_p \otimes {\bf u}_p \rangle}{k_p} + (1-f_s) \frac{2}{3} {\bf I} \right] 
\end{equation}
where $\varepsilon_p$ is one-half the trace of $\bds{\varepsilon}_p$.

\subsection{Statistically homogeneous particle-phase model}\label{sec::shppm}

For statistically homogeneous flow, the Eulerian equations corresponding to the stochastic equation system \eqref{eq:SDEp-homo-coll-first}--\eqref{eq:SDEp-homo-coll-fourth} are the following (see Part~I for their derivation):
\begin{equation}
\label{eq:feq_Up-hom}
\frac{d \lra{{\bf U}_{p}}}{d t} =
 \frac{1}{\tau_p} \langle {\bf U}_s - {\bf U}_p \rangle  + {\bf g},
\end{equation}
\begin{equation}
\langle \bds{\delta} {\bf v}_p \rangle = 0,
\end{equation}
\begin{equation}
\label{eq:feq_Us-hom}
\frac{d \lra{{\bf U}_{s}}}{d t}  =
- \frac{1}{\rho_f}  \nabla \langle p_f \rangle +  {\bf G} \bcdot (\langle {\bf U}_s \rangle - \langle {\bf U}_f \rangle) + \frac{\varphi}{\tau_p}   \langle {\bf U}_p - {\bf U}_s \rangle + {\bf g},
\end{equation}
The particle-phase pressure tensor, $\langle {\bf P} \rangle = \langle \delta {\bf v}_p \otimes \delta {\bf v}_p \rangle$, is found from
\begin{equation}
\frac{d \langle {\bf P} \rangle}{d t} = \bds{\mathcal{P}}_{P}  + \bds{\varepsilon}_p - \frac{2}{\tau_p} \langle {\bf P} \rangle  + \frac{1}{2 \tau_c} [ (1+e)^2 \langle \Theta_p \rangle {\bf I} - (1+e)(3-e) \langle {\bf P} \rangle ] 
\end{equation}
where $\langle \Theta_p \rangle = \frac{1}{3} trace( \langle {\bf P} \rangle )$ and the production term due to mean velocity gradients is 
\begin{equation}
\bds{\mathcal{P}}_{P} = - (\langle {\bf P} \rangle \bcdot \bfnabla \langle {\bf U}_{p}\rangle )^{\dagger} 
\end{equation}
where  the symbol $(\cdot)^{\dagger}$ implies the summation of a second-order tensor with its transpose.

For the particle-phase Reynolds-stress tensor, we obtain
\begin{equation}
\label{eq:feq_upup-hom}
\frac{d \lra{{\bf u}_{p} \otimes {\bf u}_{p}}}{d t}  =
\bds{\mathcal{P}}_p + \bds{\mathcal{R}}_p - \bds{\varepsilon}_p .
\end{equation}
The redistribution term is expressed as
\begin{equation}\label{eq:LLR-IP}
\bds{\mathcal{R}}_p = - C_{Rp} \frac{\varepsilon_p}{k_p} \left( \langle {\bf u}_p \otimes {\bf u}_p \rangle  
- \frac{2}{3} k_p {\bf I} \right) 
\end{equation}
with $k_{p} = \frac{1}{2} \lra{{\bf u}_{p} \bcdot {\bf u}_{p}}$ and 
\begin{equation}
C_{Rp} = 1 + \frac{3}{2} C_{0p} 
\end{equation}
where 
$C_{0p}$ is the model constant in \eqref{eq:SDEp-homo-coll-second}.
The production term in \eqref{eq:feq_upup-hom} is defined by $\bds{\mathcal{P}}_{p}=\bds{\mathcal{P}}_{Sp}+\bds{\mathcal{P}}_{Dp}$ where
$\bds{\mathcal{P}}_{Sp}$ is the mean-shear-production term, given by
\begin{equation}
\bds{\mathcal{P}}_{Sp} = - (\lra{{\bf u}_{p} \otimes {\bf u}_{p}} \bcdot \bfnabla \langle {\bf U}_{p}\rangle )^{\dagger} ;
\end{equation}
and $\bds{\mathcal{P}}_{Dp}$ is the drag-production term, given by
\begin{equation}
\bds{\mathcal{P}}_{Dp} = \frac{1}{\tau_p}(\lra{ {\bf u}_s \otimes {\bf u}_{p} }^{\dagger} - 2 \langle {\bf u}_p \otimes {\bf u}_p \rangle ) .
\end{equation}

The fluid-seen Reynolds-stress tensor is found from
\begin{equation}
\label{eq:feq_usus}
\frac{d \lra{{\bf u}_{s} \otimes {\bf u}_{s}} }{d t} = \bds{\mathcal{P}}_{s} +
{ ( {\bf G}} \bcdot \lra{ {\bf u}_{s} \otimes {\bf u}_{s} } )^{\dagger}
+ \lra{{\bf B}_s {\bf B}_s^T} 
\end{equation}
where ${\bf B}_s$ is the diffusion matrix in \eqref{eq:SDEp-homo-coll-fourth} and $k_{f@p} = \frac{1}{2} \langle ({\bf U}_s - \langle {\bf U}_f \rangle) \cdot ({\bf U}_s - \langle {\bf U}_f \rangle) \rangle $.
The production term in \eqref{eq:feq_usus} is defined by $\bds{\mathcal{P}}_{s}=\bds{\mathcal{P}}_{Ss}+\bds{\mathcal{P}}_{Ds}$ where
$\bds{\mathcal{P}}_{Ss}$ is the mean-shear-production term, given by
\begin{equation}
\bds{\mathcal{P}}_{Ss} = - ( \lra{{\bf u}_{s} \otimes {\bf u}_{s}} \bcdot \bfnabla \langle {\bf U}_{s}\rangle )^\dagger 
;
\end{equation}
and $\bds{\mathcal{P}}_{Ds}$ is the drag-production term, given by
\begin{equation}
\bds{\mathcal{P}}_{Ds} = 
\frac{ \varphi}{\tau_p} ( \lra{{\bf u}_s \otimes {\bf u}_p}^{\dagger} - 2 \lra{{\bf u}_{s} \otimes {\bf u}_{s}} ).
\end{equation}

The fluid--particle covariance Reynolds-stress tensor is found from
\begin{equation}
\label{eq:feq_upus-hom}
\frac{d \lra{{\bf u}_{s} \otimes {\bf u}_{p}} }{d t} = \bds{\mathcal{P}}_{sp} +
{  {\bf G}} \bcdot \lra{ {\bf u}_{s} \otimes {\bf u}_{p} }^{T} - \frac{1}{{T}_{Lp}} \lra{ {\bf u}_{p} \otimes {\bf u}_{s} } 
\end{equation}
where $k_{fp} = \frac{1}{2} \lra{{\bf u}_{s} \bcdot {\bf u}_{p}}$.
The production term in \eqref{eq:feq_upus-hom} is defined by $\bds{\mathcal{P}}_{sp}=\bds{\mathcal{P}}_{Ssp}+\bds{\mathcal{P}}_{Dsp}$ where
$\bds{\mathcal{P}}_{Ssp}$ is the mean-shear-production term, given by
\begin{equation}
\bds{\mathcal{P}}_{Ssp} = - \lra{{\bf u}_{s} \otimes {\bf u}_{p}} \bcdot \bfnabla \langle {\bf U}_{p}\rangle ^T
- ( \lra{{\bf u}_{p} \otimes {\bf u}_{p}} + \langle {\bf P} \rangle ) \bcdot \bfnabla \langle {\bf U}_{s}\rangle ^T
;
\end{equation}
and $\bds{\mathcal{P}}_{Dsp}$ is the drag-production term, given by
\begin{equation}
\bds{\mathcal{P}}_{Dsp} = 
\frac{1}{\tau_p} (\lra{ {\bf u}_{s} \otimes {\bf u}_{s}} - \lra{{\bf u}_p \otimes {\bf u}_s}) +
\frac{\varphi}{\tau_p} ( \lra{{\bf u}_p \otimes {\bf u}_p} - \lra{{\bf u}_{s} \otimes {\bf u}_{p}})  .
\end{equation}
The particle Lagrangian time scale, introduced in \eqref{eq:SDEp-homo-coll-second}, is defined as
\begin{equation}
T_{Lp} = \frac{2}{\left( 1 + \frac{3}{2} C_{0p} { + f_s}
	\right)} \frac{k_p}{\varepsilon_p} .
\label{eq:T_lp}
\end{equation}

The particle-phase dissipation is modelled through an Eulerian equation in analogy to single-phase flows \citep{Fox2014}:
\begin{align}
\frac{d \varepsilon_p }{d t}  =  (C_{\epsilon 1 p} \mathcal{P}_{Sp} - C_{\epsilon 2p} \varepsilon_p) \frac{\varepsilon_p}{k_p} +   \frac{C_{3p}}{\tau_p} \left( \frac{ k_{fp}}{k_{f@p}} \varepsilon_f - \beta_p \, \varepsilon_p \right)
\label{eq:epsilon_p1}
\end{align}
where $C_{\epsilon 1 p}$, $C_{\epsilon 2p}$, $C_{3p}$ and $\beta_p$ are model parameters. 
Finally, all the fluid-phase quantities are obtained through the RA equations presented in \S\ref{sec:fluid-model}.

\subsection{Simplified model for particle phase }
\label{sec:part-model-simplified}

We propose here  a simplified model for the particle phase, where collisions between particles are neglected and only the total particle velocity is modelled, thus loosing information about its decomposition into the correlated and uncorrelated parts. In particular, this corresponds to assuming that the particle velocity coincides with the correlated part, i.e.\ ${\bf V}_p = {\bf U}_p$.  With this hypothesis, we recover the model previously proposed by \cite{Min_04} and \cite{Pei_06} for the fluid velocity seen by the particles, but with a modified diffusion term, as discussed in detail in Part~I. 
The resulting set of SDEs for the simplified model is
\begin{equation} \label{eq:SDEp-homo}
\left\{\begin{split}
& d{ x}_{p,i}(t)= { V}_{p,i}\,dt, \\
& d{ V}_{p,i}(t)= \frac{{ U}_{s,i}-{ V}_{p,i}}{\tau_p}\,dt + { g}_i \,dt, \\
& d{ U}_{s,i}(t) = - \frac{1}{\rho_f} \frac{\partial \langle p_f \rangle}{\partial x_i} \, dt  - \frac{1}{{ T}_{L,i}^*} ({ U}_{s,i} - \langle { U}_{f,i} \rangle ) \,dt  - \varphi \left( \frac{{ U}_{s,i} - { V}_{p,i}}{\tau_p} \right) dt   + { g}_i \, dt \\ & \qquad \qquad + \Bigl [ \varepsilon_f \Bigl ( C_{0f} {b}_i \frac{\tilde{k}_f}{k_f} + \frac{2}{3} \Bigl ( {b}_i \frac{\tilde{k}_f}{k_f} -1 \Bigl) \Bigl )  +  2 \varphi  \frac{\langle { V}_{p,i} - { U}_{s,i} \rangle}{\tau_p} (\langle { U}_{s,i} \rangle - \langle {U}_{f,i} \rangle) \\ & \qquad \qquad  {\qquad - 2 \frac{\langle \alpha_p \rangle}{\langle \alpha_f \rangle \rho_f}\frac{\partial \langle p_f \rangle}{\partial x_i} (\langle U_{s,i} \rangle - \langle U_{f,i}\rangle)} \Bigl ]^{{1}/{2}}  \, d{ W}_{s,i} 
\end{split}\right.
\end{equation}
where all of the parameters were defined in the complete model above.

The corresponding Eulerian RA equations for statistically homogeneous flow are
\begin{equation}
\label{eq:feq_Up-hom2}
\frac{d \lra{{\bf V}_{p}}}{d t} =
 \frac{1}{\tau_p} \langle {\bf U}_s - {\bf V}_p \rangle  + {\bf g},
\end{equation}
\begin{equation}
\label{eq:feq_Us-hom2}
\frac{d \lra{{\bf U}_{s}}}{d t}  =
- \frac{1}{\rho_f}  \nabla \langle p_f \rangle - \frac{1}{{\bf T}_L^*} \circ (\langle {\bf U}_s \rangle - \langle {\bf U}_f \rangle) + \frac{\varphi}{\tau_p}   \langle {\bf V}_p - {\bf U}_s \rangle + {\bf g} .
\end{equation}
For the second-order moments, we obtain
\begin{equation}
\label{eq:feq_upup-hom2}
\frac{d \lra{{\bf v}_{p} \otimes {\bf v}_{p}}}{d t}  = 
\bds{\mathcal{P}}_{Vp}
{ +}
\frac{1}{\tau_p}(\lra{ {\bf u}_s \otimes {\bf v}_{p} } ^{\dagger} - 2 \langle {\bf v}_p \otimes {\bf v}_p \rangle ),
\end{equation}
\begin{equation}
\label{eq:feq_upus-hom2}
\begin{split}
\frac{d \lra{{\bf u}_{s} \otimes {\bf v}_{p}}}{d t}  = \bds{\mathcal{P}}_{Vsp} -
\frac{1}{{\bf T}_L^*} \circ \lra{ {\bf u}_{s} \otimes {\bf v}_{p} } 
+\frac{1}{\tau_p} (\lra{ {\bf u}_{s} \otimes {\bf u}_{s}} - \lra{{\bf v}_{p} \otimes {\bf u}_{s}})  \\
+ \frac{\varphi}{\tau_p} ( \lra{{\bf v}_p \otimes {\bf v}_p} - \lra{{\bf u}_{s} \otimes {\bf v}_{p}}),
\end{split}
\end{equation}
\begin{equation}
\label{eq:feq_usus2}
\frac{d \lra{{\bf u}_{s} \otimes {\bf u}_{s}}}{d t}  = \bds{\mathcal{P}}_{Ss} -
\frac{2}{{\bf T}_L^*} \circ \lra{ {\bf u}_{s} \otimes {\bf u}_{s} } 
+ \lra{{\bf B}_s {\bf B}_s^T} 
+ \frac{\varphi}{\tau_p} ( \lra{{\bf u}_s \otimes {\bf v}_p}^{\dagger} - 2 \lra{{\bf u}_{s} \otimes {\bf u}_{s}} )
\end{equation}
The mean-shear-production terms are
\begin{equation}
\bds{\mathcal{P}}_{Vp} = 
- (\lra{{\bf v}_{p} \otimes {\bf v}_{p}} \bcdot \bfnabla \langle {\bf V}_{p}\rangle )^{\dagger}
\end{equation}
and
\begin{equation}
\bds{\mathcal{P}}_{Vsp} = 
- \lra{{\bf u}_{s} \otimes {\bf v}_{p}} \bcdot \bfnabla \langle {\bf V}_{p}\rangle ^T
- \lra{{\bf v}_{p} \otimes {\bf v}_{p}} \bcdot \bfnabla \langle {\bf U}_{s}\rangle ^T.
\end{equation}

\subsection{Fluid-phase model}
\label{sec:fluid-model}

The Eulerian RA equation describing the fluid phase mass balance for a statistically homogeneous flow reduces to
\begin{equation}
\frac{d \langle \alpha_f \rangle}{d t} = 0
\label{eq:alfa_f}
\end{equation}
i.e., $\langle \alpha_f \rangle$ is constant. The fluid-phase velocity and Reynolds stresses are found from
\begin{equation}
 \frac{d \langle {\bf U}_{f} \rangle}{d t} 
 = - \frac{1}{\rho_f \langle \alpha_f \rangle} \bfnabla \langle p_f \rangle + \frac{\varphi}{\tau_p}   \langle {\bf U}_{p}  -  {\bf U}_{s} \rangle  + {\bf g} ,
\label{eq:momentum-homo}
\end{equation}
and
\begin{equation}
\frac{d \langle {\bf u}_f \otimes {\bf u}_f \rangle}{d t} 
= \bds{\mathcal{P}}_f  - C_{Rf} \frac{\varepsilon_f}{k_f} \left( \langle {\bf u}_f \otimes {\bf u}_f \rangle - \frac{2}{3} k_f {\bf I} \right) - C_{2f} \left( {\bds{\mathcal{P}}_{Sf} - \frac{2}{3} {\mathcal{P}}_{Sf} {\bf I}} \right)  - \frac{2}{3} \varepsilon_f {\bf I}
\label{tkef:homo}
\end{equation} 
where $k_{f} = \frac{1}{2} \lra{{\bf u}_{f} \bcdot {\bf u}_{f}}$. The  production term is $\bds{\mathcal{P}}_f = \bds{\mathcal{P}}_{Sf} + \bds{\mathcal{P}}_{Df}$ where  $\bds{\mathcal{P}}_{Sf}$ is the mean-shear-production term, given by
\begin{equation}
\bds{\mathcal{P}}_{Sf} = - (\lra{{\bf u}_{f} \otimes {\bf u}_{f}} \bcdot \bfnabla \langle {\bf U}_{f}\rangle )^{\dagger} ;
\end{equation}
and $\bds{\mathcal{P}}_{Df}$ is the drag-production term, given by
\begin{equation}
\bds{\mathcal{P}}_{Df} = \frac{\varphi}{\tau_p}  [ \langle {\bf u}_s \otimes ({\bf u}_p - {\bf u}_s) \rangle + \langle {\bf U}_s   -  {\bf U}_f \rangle \otimes  {\langle {\bf U}_p - {\bf U}_s \rangle}  ]^{\dagger} .
\end{equation}
In \eqref{tkef:homo}, $C_{Rf}$ is the Rotta constant for the redistribution \citep{Pope_turbulent}, 
and $\mathcal{P}_{Sf} = \frac{1}{2} trace (\bds{\mathcal{P}}_{Sf})$. 

The fluid-phase dissipation equation is 
\begin{equation}
 \frac{d \varepsilon_f }{d t} =  \left( C_{\epsilon 1 f} \mathcal{P}_{Sf} -  C_{\epsilon 2 f}   \varepsilon_f \right) \frac{\varepsilon_f}{k_f}  +  C_{3f} \frac{\varphi}{\tau_p} \left(  \frac{k_{fp}}{k_{f@p}} \varepsilon_p - \beta_f \varepsilon_f \right)  + C_4 \frac{\varepsilon_p}{k_p} \mathcal{P}_{D}
 \label{eq:epsilon2-homo}
\end{equation}
where $C_{\epsilon 1 f}$, $C_{\epsilon 2 f}$, $C_{3f}$, $\beta_f$ and $C_4$ are model constants,  and 
\begin{equation}\label{eq:PDF}
\mathcal{P}_{D} = \frac{\varphi}{\tau_p} {\frac{
\langle {\bf U}_s - {\bf U}_f \rangle \bcdot \langle {\bf U}_p - {\bf U}_f \rangle}{2}}.
\end{equation}

If the RA equations for the fluid phase are coupled with the simplified model described in \S\ref{sec:part-model-simplified}, ${\bf U}_p$ must be replaced with ${\bf V}_p$. Moreover, the particle-phase Lagrangian time-scale $k_p / \varepsilon_p$ is not specified, and it is thus replaced by a fluid time-scale through a proportionality constraint:
\begin{equation} 
\frac{\varepsilon_p}{k_p} = \alpha \frac{\varepsilon_f}{k_{f@p}}
\label{eq:time-scales}
\end{equation}
Now, substituting (\ref{eq:time-scales}) in (\ref{eq:epsilon2-homo}) and incorporating $\alpha$ in the model constants, gives the following equation for dissipation:
\begin{equation}
\frac{d \varepsilon_f }{d t} =  \left( C_{\epsilon 1 f} \mathcal{P}_{Sf} -  C_{\epsilon 2 f}   \varepsilon_f \right) \frac{\varepsilon_f}{k_f}  
+  C_{3f} \frac{\varphi}{\tau_p} \left(  \frac{ k_p}{k_{f@p}} k_{fp}   
-  \beta_f k_{f@p} \right) \frac{\varepsilon_f}{k_{f@p}}  + C_4 \frac{\varepsilon_f}{k_{f@p}} \mathcal{P}_{D}
 \label{eq:epsilon2b-homo}
\end{equation}
The values of $C_{3f}$ and $C_4$ in \eqref{eq:epsilon2b-homo} may need to be adjusted as compared to \eqref{eq:epsilon2-homo} to account for the alternative time scale.

\section{Numerical Results}\label{sec::NR}

If the fluid--particle flow is spatially homogeneous as in the cases that we are going to test below, the equations can be simplified, since hydrodynamic variables are invariant in space (see \S\ref{sec::shppm}). Moreover the Eulerian RA equations for the particle phase obtained from the set of Lagrangian stochastic equations, in this case are in closed form (see \S\ref{sec::shppm}). Although not needed here, the great advantage of the Lagrangian form is that it can be applied also to inhomogeneous flows without needing any additional considerations. 

We present three spatially homogeneous examples of increasing complexity: (i) isotropic turbulence with one-way coupling \citep{fevrier2005partitioning}, (ii) isotropic decaying \citep{sundaram1999numerical} and sheared turbulence \citep{ahmed2000mechanisms} with two-way coupling, and (iii) gravity-driven CIT \citep{Fox2015}. The first example is aimed at appraising the partitioning of the particle kinetic energy, the second at testing the dynamics in the absence/presence of shear production (i.e., $\bds{\mathcal{P}}_{Sf}$) without a mean velocity difference, and the third at validating the model for production due to a mean velocity difference (i.e., \eqref{eq:PDF}).

\subsection{Homogeneous isotropic turbulence}

In order to illustrate the effectiveness of the decomposition of the particle velocity, we apply the models developed for the particle phase to the homogeneous isotropic turbulence simulations of \cite{fevrier2005partitioning} for non-collisional particles. For this example, the mean velocities ${\bf U}_p, \, {\bf U}_f, \, {\bf U}_s$ are null, and $\varphi = 0$. At a first glance it may appear odd to compare the results of a model developed for collisional flows to DNS data for non-collisional particles. However, the crucial point for the modelling is the correlation between the fluid and particle velocities as captured by $k_f$ and $k_p$, respectively. The applicability of the proposed models to non-collisional flows depends on the model used for turbulent dissipation, since the relative balance between $k_p$ and $\Theta_p$ is determined by $\varepsilon_p$, for both dilute and dense flows. The scope of this section is thus to verify if in the dilute case, where collisions do not play any role, energy budgets are well predicted. 

The mesoscale DNS simulations of \cite{fevrier2005partitioning} use one-way coupling with stationary fluid turbulence, and a particle-Reynolds-number-dependent drag coefficient $f_D$ (instead of a constant $\tau_p$). Therefore, the drag time scale is Stokes-number-dependent, and only qualitative comparisons can be made. 

When a cloud of particles is put into a box filled with a homogeneous and isotropic turbulent flow and is being agitated by the fluid turbulence, then, after a transient period, the statistics of particle velocities 
reach equilibrium values. These limit values are of course functions
of the (constant) statistics of the fluid (its mean kinetic energy,
the Lagrangian timescale, among others). The relations giving
the equilibrium values in terms of the fluid statistics are called
the Tchen's relations. They were first obtained by \cite{tchen1947mean}
and later reformulated by \cite{Hin_75}. In Tchen or Hinze's works, the determination of the equilibrium values was obtained through spectral analysis and manipulation of the fluid and particle energy spectra, where the fluid spectrum is assumed to have an exponential form. This derivation can be cumbersome and the physical meaning of the exponential form is not obvious. On the other hand, the same relations are derived from the Lagrangian pdf model in a straightforward way. 

In forced, homogeneous, isotropic turbulence without body forces, all mean velocities are zero and the Reynolds-stress and particle-phase pressure tensors are isotropic. Moreover, $k_{f@p} = k_f$. With one-way coupling and fixed $k_f$ and $\varepsilon_f$, the relevant moment equations from the complete model for the particle phase reduce to
\begin{subequations}\label{eq:hitcomplete}
\begin{gather}
\frac{d k_p}{d t}  = \frac{2}{\tau_p}( k_{fp} - k_p ) - \varepsilon_p , \label{eq:hitk_p} \\
\frac{3}{2} \frac{d \langle \Theta_p \rangle}{d t} = - \frac{3 \langle \Theta_p  \rangle }{\tau_p} + \varepsilon_p , \label{eq:hitT_p} \\
\frac{d k_{fp}}{d t}  = - \left( \frac{1}{T_{Lf}} + \frac{1}{T_{Lp}} \right) k_{fp} + \frac{1}{\tau_p}( k_{f} - k_{fp} )  , \label{eq:hitk_fp} \\
 \frac{d \varepsilon_p }{d t}  = - C_{\epsilon 2p}  \frac{\varepsilon_p^2}{k_p} +  \frac{C_{3p}}{ \tau_p} \left( \frac{ k_{fp}}{k_{f}} \varepsilon_f - \beta_p  \varepsilon_p \right) 
 \label{eq:hitepsilon_p}
\end{gather}
\end{subequations}
where $T_{Lf}$ is given by \eqref{eq:T_L} and $T_{Lp}$ by \eqref{eq:T_lp}.
After a transient period, all the statistics reach their steady-state values.
This yields 
\begin{subequations}
\begin{gather}
2 \frac{k_{fp} - k_p}{\tau_p} - \varepsilon_p = 0, \label{eq:up_tchen}\\
- \frac{3\langle \Theta_p \rangle}{\tau_p} + \varepsilon_p  =0, \label{eq:dvp_tchen}\\  
\frac{k_{f}}{\tau_p} - \left( \frac{1}{\tau_p} + \frac{1}{T_{Lf}} + \frac{1}{T_{Lp}} \right) k_{fp}=0, \label{eq:upus_tchen} \\
St_p^2 -  \frac{C_{3p}}{C_{\epsilon 2p}} \left( \frac{ k_{fp}}{k_p} St_f - \beta_p  St_p \right) = 0
\label{eq:sshitepsilon_p}
\end{gather}
\end{subequations}
where $St_p = \tau_p {\varepsilon_p}/{k_p}$ and $St_f = \tau_p {\varepsilon_f}/{k_f}$.
Summing \eqref{eq:up_tchen} and \eqref{eq:dvp_tchen} then yields
\begin{equation}
k_{fp} = \kappa_p ,
\end{equation}
which can be used together with \eqref{eq:upus_tchen} to obtain a Tchen-like relation: 
\begin{equation}
\kappa_p = \frac{1}{1+ \tau_p/T_L'} k_{f}
\label{eq:tchen}
\end{equation}
with $C_0 = C_{0p} = C_{0f}$ and
\begin{equation}
\frac{1}{T_L'} =  \frac{1}{T_{Lf}} + \frac{1}{T_{Lp}} 
= \left( \frac{1}{2} + \frac{3}{4} C_0 \right) \left( \frac{\varepsilon_f}{k_f} + \frac{\varepsilon_p}{k_p} \right).
\end{equation}
Here, $\tau_p/T_L'$ is an effective integral-scale Stokes number for the particles.  Furthermore, $St_p$ is constant, and can be related to $St_f$ using \eqref{eq:up_tchen} and \eqref{eq:sshitepsilon_p}. With $\beta_p=1$, this relation depends only on the parameter ratio $\frac{C_{3p}}{C_{\epsilon 2p}}$, and thus $St_p=St_f$ when $C_{3p}=2C_{\epsilon 2p}$.  Note that the value of $St_p$ controls the ratio $k_p / \kappa_p = 2/(2 + St_p)$ and, as expected, all of the particle-phase kinetic energy is spatially correlated when $St_p=0$.

\begin{figure}
\center
\begin{tabular}{cc}
{\includegraphics[width=.44\textwidth]{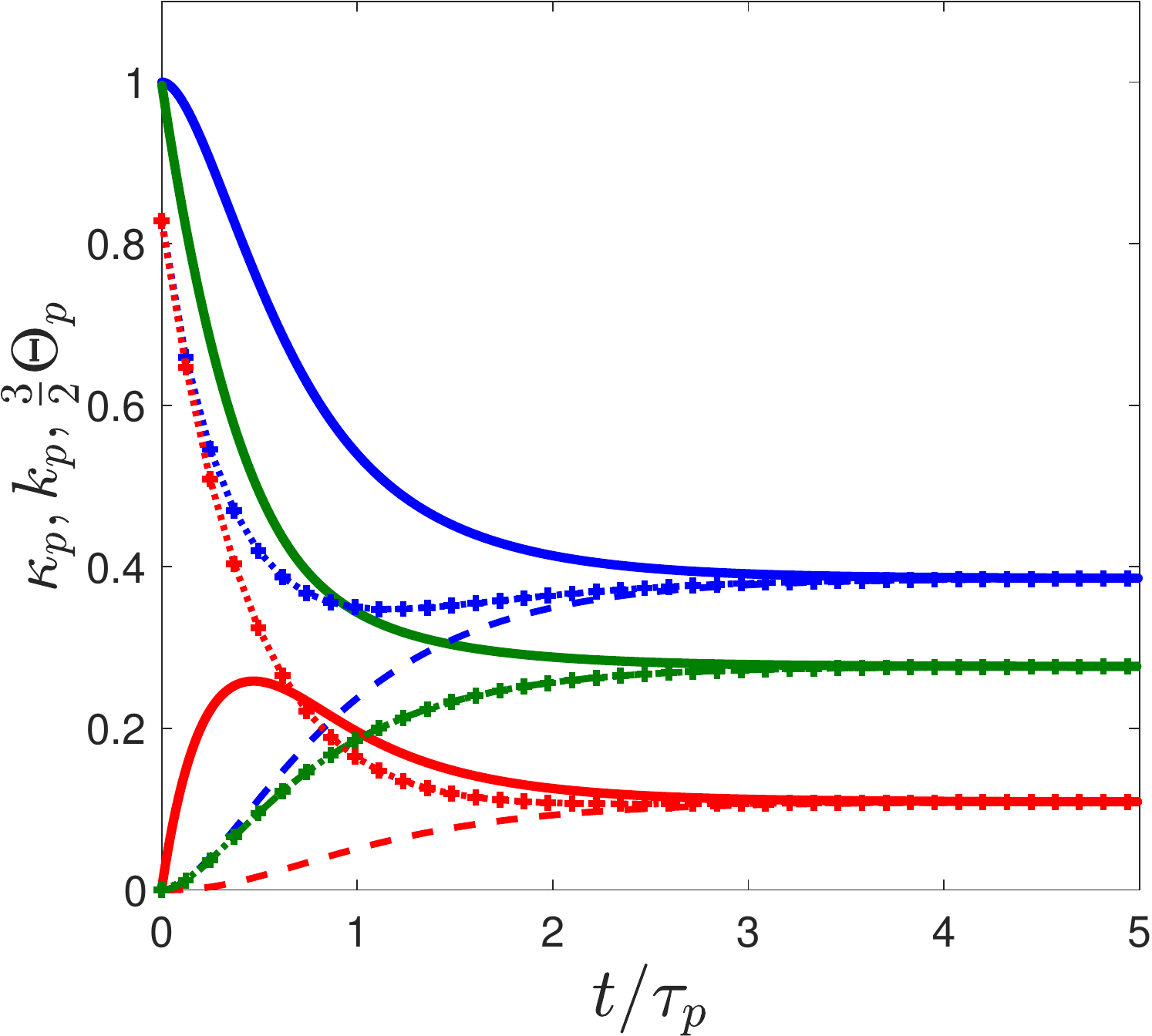}}
& {\includegraphics[width=.43\textwidth]{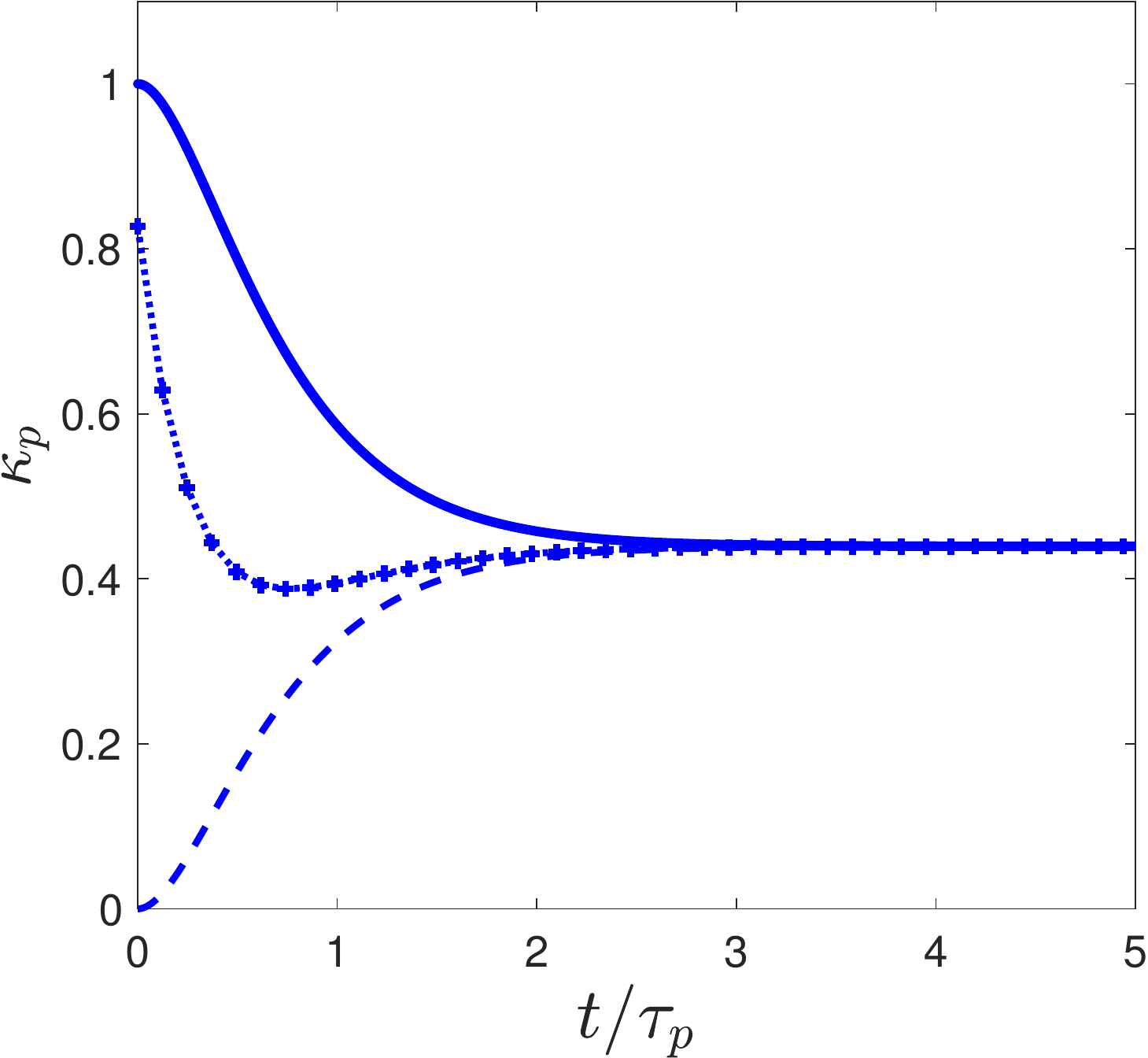}}\\
\hspace{0.5cm}(a) & \hspace{0.5cm}(b)\\ \\
{\includegraphics[width=.45\textwidth]{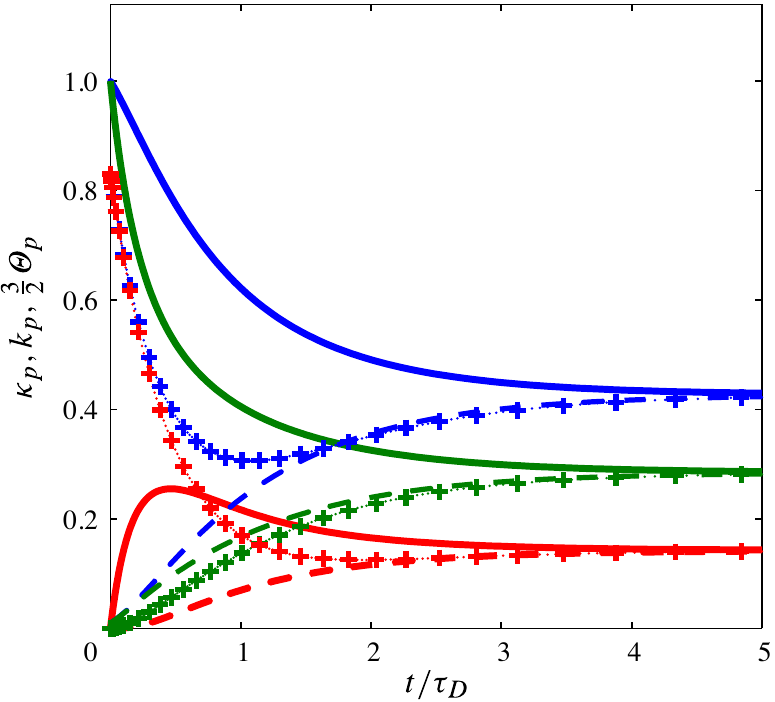}}
 & {\includegraphics[width=.45\textwidth]{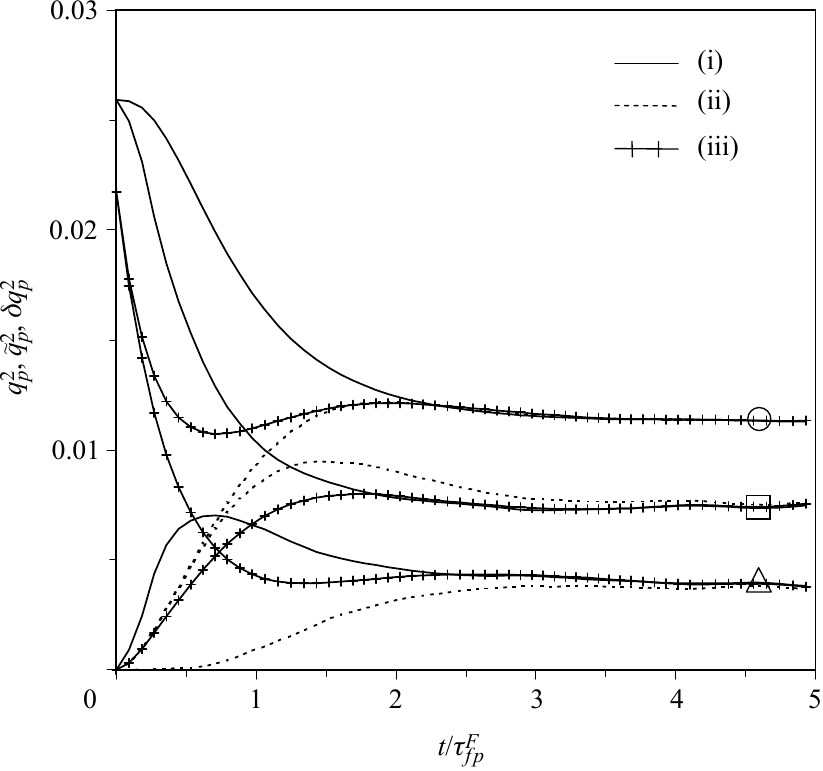}}\\
\hspace{0.5cm}(c) & \hspace{0.7cm}(d)
\end{tabular}
\caption{Time evolution of the dimensionless particle-phase energy components for $St_f=0.81$. The curves correspond to the three considered sets of initial conditions:  case (i) solid lines; (ii) dashed lines; and (iii) $+$ line. $\kappa_p$ is plotted in blue, $k_p$ in green and $\frac{3}{2} \Theta_p$ in red.
(a) Simulations with the complete particle model \eqref{eq:hitcomplete}.
(b) Simulations carried out with the simplified particle model \eqref{eq:SDEp-homo} with $C_0 =2.1$. Only total kinetic energy $\kappa_p$ is computed. 
(c) Results from \cite{Fox2014} with the same initial conditions as in (a).
(d) Results of point-particle DNS from figure 8 of \cite{fevrier2005partitioning}, where the results are in dimensional form. The notation in \cite{fevrier2005partitioning} is $q_p^2 = \kappa_p$, $\widetilde{q}_p^2 = k_p$, $\delta q_p^2 = \frac{3}{2} \Theta_p$ and $\tau_{fp}^F \propto \tau_D$.}
\label{Fig:tchen}
\end{figure}

\cite{fevrier2005partitioning} presented time-dependent DNS results of particle-laden homogeneous and isotropic turbulence for $St_f = 0.81$ and $\varphi=0$, for three sets of initial conditions: (i) $\kappa_p = k_{fp} = 1$, $k_p = 1$; (ii) $\kappa_p = k_{fp} = 0$, $k_p = 0$; and (iii) $\kappa_p = 0.83$, $k_p = k_{fp} = 0$.  We reproduced the same cases by solving the dimensionless forms of system \eqref{eq:hitcomplete} with the following values of the model constants: $C_0=1$, $C_{\epsilon 2p} = 1.92$, $C_{3p} = 3.5$ and $\beta_p=1$. For consistency with $k_p$, $\varepsilon_p$ is initially set to zero when $k_p = 0$ and for case (i) the initial value of dissipation is $\varepsilon_p =2$. Figure \ref{Fig:tchen}(a) shows the time evolution of $\kappa_p$, $k_p$ and $\Theta_p$ obtained with the present model for the three different sets of considered initial conditions, while the evolution of $\kappa_p$ obtained with the simplified model for the same cases is reported in figure \ref{Fig:tchen}(b). Moreover, \ref{Fig:tchen}(c) shows the same quantities as in figure \ref{Fig:tchen}(a) obtained with the model proposed in \cite{Fox2014} and  \ref{Fig:tchen}(d) the results of the DNS of \cite{fevrier2005partitioning}. In all cases, after a transient a steady state is reached, as expected. It can be seen how in DNS the total particle kinetic energy is distributed in the correlated part and in the uncorrelated granular temperature. This energy partition is satisfactorily captured by our complete model as well as by the model proposed in \cite{Fox2014}. Clearly, the simplified model can only give the total energy $\kappa_p$, which is however in good agreement with that of DNS and of more complete models. The transient behaviour is also in very good qualitative agreement with that obtained in DNS.

\subsection{Decaying and homogeneous-shear flow}\label{sec:dhsf}

In this section we focus on the particle--turbulence interactions in homogeneous flows, and, in particular, on the cases simulated by DNS in \cite{sundaram1999numerical} for decaying turbulence and in \cite{ahmed2000mechanisms} for homogeneous-shear flows. For these examples, the mean velocities ${\bf U}_p, \, {\bf U}_f, \, {\bf U}_s$ are null. In both cases the particle-phase volume fraction is such that two-way interactions need to be considered: flow modification by non-collisional point particles reveal a non-trivial dependence on the particle Stokes number and mass loading $\varphi$. It is thus interesting to verify if our model is able to reproduce such physics and if the same dependencies on the particle Stokes number are found.

The Eulerian RA equations describing the fluid and particle phases for the considered cases are summarized below for the sake of completeness. In these equations, the flow is statistically homogeneous with a constant shear $\mathcal{S}_f = \partial \langle U_{f,1} \rangle / \partial x_2$ ($\mathcal{S}_f = 0$ in decaying turbulence) and with gravity and collisions neglected. The non-zero components of the second-order moments are $(i,j) = (1,1), (1,2), (2,1), (2,2), (3,3)$. 
For the fluid phase, the Reynolds stresses are found from
\begin{subequations}\label{eq:hitcompletef}
\begin{multline}
\frac{d \langle { u}_{f,i} { u}_{f,j} \rangle}{d t} = \mathcal{P}_{f,ij}  
- C_{Rf} \frac{\varepsilon_f}{k_f} \left( \langle { u}_{f,i} { u}_{f,j} \rangle - \frac{2}{3} k_f \delta_{ij} \right)
- C_{2f} \left( { \mathcal{P}_{Sf,ij} - \frac{2}{3} \mathcal{P}_{Sf} \delta_{ij}} \right)  \\
- \frac{2}{3} \varepsilon_f \delta_{ij} ,
\end{multline}
\begin{equation}
 \frac{d \varepsilon_f }{d t} = (C_{\epsilon 1 f} \mathcal{P}_{Sf} -  C_{\epsilon 2 f}\varepsilon_f) \frac{\varepsilon_f}{k_f} +  C_{3f} \frac{\varphi }{\tau_p} \left( \frac{k_{fp}}{k_{f@p}}  \varepsilon_p - \beta_f \varepsilon_f \right)
\end{equation} 
\end{subequations}
with production terms due to mean shear and drag:
\begin{subequations}\label{eq:hitcompletePf}
\begin{gather}
\mathcal{P}_{f,ij} = \mathcal{P}_{Sf,ij} + \mathcal{P}_{Df,ij} , \\
\mathcal{P}_{Sf,ij} = - \langle u_{f,i} u_{f,2} \rangle \mathcal{S}_f \delta_{1j} - \langle u_{f,j} u_{f,2} \rangle \mathcal{S}_f \delta_{1i} ,
\\
\mathcal{P}_{Df,ij} = \frac{\varphi}{\tau_p}  ( \langle { u}_{s,i} u_{p,j} \rangle + \langle { u}_{s,j} u_{p,i} \rangle - 2 \langle u_{s,i} u_{s,j} \rangle ) .
\end{gather}
\end{subequations}
For the particle phase, the pressure tensor and Reynolds stresses are found from
\begin{subequations}\label{eq:hitcompletep}
\begin{equation}
\frac{d \langle P_{ij} \rangle}{d t} = \mathcal{P}_{P,ij} +
\varepsilon_p \left[ f_s \frac{ \langle u_{p,i}  u_{p,j} \rangle}{k_p} + (1-f_s) \frac{2}{3} \delta_{ij} \right] 
- \frac{2}{\tau_p} \langle P_{ij} \rangle ,
\end{equation}
\begin{multline}
\frac{d \lra{{u}_{p,i} {u}_{p,j}}}{d t}  =
 \mathcal{P}_{p,ij} 
 - C_{Rp} \frac{\varepsilon_p}{k_p} \left( \langle {u}_{p,i}{u}_{p,j} \rangle  - \frac{2}{3} k_p \delta_{ij} \right)  
  \\
 - \varepsilon_p \left[ f_s \frac{ \langle u_{p,i}  u_{p,j} \rangle}{k_p} + (1-f_s) \frac{2}{3} \delta_{ij} \right]  ,
\end{multline}
\begin{equation}
\frac{d \varepsilon_p }{d t}  = (C_{\epsilon 1 p} \mathcal{P}_{Sp} - C_{\epsilon 2 p} \varepsilon_p) \frac{\varepsilon_p}{k_p} +  \frac{C_{3p}}{ \tau_p} \left( \frac{ k_{fp}}{k_{f@p}} \varepsilon_f - \beta_p \, \varepsilon_p \right) 
\end{equation}
\end{subequations}
with production terms:
\begin{subequations}\label{eq:hitcompletePp}
	\begin{gather}
	\mathcal{P}_{P,ij} = - \langle P_{i2} \rangle \mathcal{S}_p \delta_{1j} - \langle P_{2j} \rangle \mathcal{S}_p \delta_{1i} ,
	\\
	\mathcal{P}_{p,ij} = \mathcal{P}_{Sp,ij} + \mathcal{P}_{Dp,ij} , \\
	\mathcal{P}_{Sp,ij} = - \langle u_{p,i} u_{p,2} \rangle \mathcal{S}_p \delta_{1j} - \langle u_{p,j} u_{p,2} \rangle \mathcal{S}_p \delta_{1i} ,
	\\
	\mathcal{P}_{Dp,ij} = \frac{1}{\tau_p}( \lra{ {u}_{s,i} {u}_{p,j} } + \lra{ {u}_{s,j} {u}_{p,i} } - 2 \langle {u}_{p,i} {u}_{p,j} \rangle )  .
	\end{gather}
\end{subequations}
Note that when $\mathcal{S}_f$ is null (i.e., decaying turbulence), all second-order tensors will be isotropic so that only their traces are needed.

The mean gradients for the particle phase and fluid seen obey
\begin{subequations}\label{eq:hitcompleteS}
\begin{gather}
\frac{d \mathcal{S}_{p}}{d t} = \frac{1}{\tau_p} ( \mathcal{S}_s - \mathcal{S}_p) ,
\\
\frac{d \mathcal{S}_{s}}{d t} = \frac{1}{\tau_p} ( \mathcal{S}_f - \mathcal{S}_s) .
\end{gather}
\end{subequations}
In the following, the particle-phase velocity is initially the same as the fluid-phase velocity such that $\mathcal{S}_p(t) = \mathcal{S}_s(t) = \mathcal{S}_f$ and thus system \eqref{eq:hitcompleteS} is not needed.

The Reynolds stresses involving the fluid seen by the particles are found from
\begin{subequations}\label{eq:hitcompletes}
\begin{gather}
\frac{d \lra{{u}_{s,i} {u}_{p,j}} }{d t} = \mathcal{P}_{sp,ij} + \left(
{ G_{ij}} - \frac{1}{{T}_{Lp}} \right) \lra{ {u}_{s,i} {u}_{p,j} }  , \\
\frac{d \lra{{u}_{s,i}{u}_{s,j}} }{d t} = \mathcal{P}_{s,ij} + 
{2 G_{ij}}  \lra{ {u}_{s,i}{u}_{s,j} }
+ \varepsilon_f \left[ C_{0f} \frac{k_{f@p}}{k_f} + \frac{2}{3} \left( \frac{k_{f@p}}{k_f} - 1 \right) \right] \delta_{ij} 
\end{gather}
\end{subequations}
with production terms due to mean shear and drag:
\begin{subequations}\label{eq:hitcompletePs}
\begin{gather}
\mathcal{P}_{s,ij} = \mathcal{P}_{Ss,ij} + \mathcal{P}_{Ds,ij} , \\
\mathcal{P}_{Ss,ij} = - \langle u_{s,i} u_{s,2} \rangle \mathcal{S}_s \delta_{1j} - \langle u_{s,j} u_{s,2} \rangle \mathcal{S}_s \delta_{1i} , \\
\mathcal{P}_{Ds,ij} = \frac{\varphi}{\tau_p} 
\left( \lra{{u}_{s,i} {u}_{p,j}} + \lra{{u}_{s,j}{u}_{p,i}} 
- 2 \lra{ {u}_{s,i}{u}_{s,j} } \right) ,
\end{gather}
\end{subequations}
and
\begin{subequations}\label{eq:hitcompletePsp}
\begin{equation}
\mathcal{P}_{sp,ij} = \mathcal{P}_{Ssp,ij} + \mathcal{P}_{Dsp,ij} ,
\end{equation}
\begin{align}
\mathcal{P}_{Ssp,ij} = 
&- ( \langle u_{s,i} u_{p,2} \rangle \mathcal{S}_p + \langle P_{i2} \rangle \mathcal{S}_s + \langle u_{p,i} u_{p,2} \rangle \mathcal{S}_s ) \delta_{1j} \notag \\ 
&- ( \langle u_{s,j} u_{p,2} \rangle \mathcal{S}_p + \langle P_{2j} \rangle \mathcal{S}_s + \langle u_{p,j} u_{p,2} \rangle \mathcal{S}_s ) \delta_{1i} ,
\end{align}
\begin{equation}
\mathcal{P}_{Dsp,ij} = \frac{1}{\tau_p} 
\left[ \lra{ {u}_{s,i} {u}_{s,j}} - \lra{ {u}_{s,j} {u}_{p,i} } + \varphi ( \lra{{u}_{p,i}{u}_{p,j}} - \lra{ {u}_{s,i} {u}_{p,j} } ) \right].
\end{equation}
\end{subequations}
In \eqref{eq:hitcompletes}, $T_{Lf}$ is given by \eqref{eq:T_L} and $T_{Lp}$ by \eqref{eq:T_lp}.

It is worth noting that, even in these simple flow conditions, the model still retains some of its features, as, for instance, the distinction between the fluid kinetic energy and the fluid--particle velocity correlation, which can be computed from Lagrangian quantities by averaging, i.e. $k_{fp} = \frac{1}{2} \langle u_{p,k} u_{s,k} \rangle$, while in Eulerian models that do not account for the fluid seen by the particles (see \cite{Fox2014}), it is modeled as $k_{fp} = (k_f k_p)^{1/2}$. Moreover, having derived our model from the one proposed for dilute flows by \cite{Pei_02}, it should be remarked that only a part of the crossing trajectory effect is taken into account, that is when there is a mean drift, and thus, a mean relative velocity between fluid and particles. This means that in the case that we are testing, the modified Lagrangian timescale equals the fluid Lagrangian timescale, $T_L^* = T_L$, for all Stokes numbers. Conversely, particle inertia should affect the Lagrangian timescale of the fluid velocity seen by the particles. In particular, if we consider the limit cases, we have two situations: particles with very low inertia, i.e. $\tau_p / T_L \ll 1$, follow almost exactly the fluid, yielding $ T_L^* = T_L$ for the fluid velocity seen. Particles with high inertia, i.e. $\tau_p/T_L \gg 1$, are nearly at a standstill with respect to the fluid and therefore, the fluid velocity seen time scale is approximately the Eulerian time scale, $T_L^* = T_E$. This inconsistency has already been pointed out by \cite{Poz_98}, and, even if it can be neglected in flows where a mean drift drives the particles, becoming secondary, here it is of crucial importance, especially if we are interested in finding the trends of the decay-rate with respect to particle inertia. 
For this reason we propose to add a Stokes dependence in $C_{\epsilon2}$ of the kind $C_{\epsilon2} = C \,(1 - \varphi St)$ with $C=1.92$, in order to retrieve the good trend with $St$. Note that this simple model, which was also used in \cite{Fox2014} for the same test case, is just qualitative and valid for the range of conditions considered herein. A more refined analysis may be necessary for general situations. 

\subsubsection{Decaying turbulence}

Concerning the values of other model constants, they are the same as in the stationary case, i.e. $C_{\epsilon2} = C_{\epsilon2p} = C_{\epsilon2f}$, $\beta = \beta_f = \beta_p =1$, $C_3 = C_{3f} = C_{3p} = 3.5$ and $C_0 = C_{0f} = C_{0p} = 1$.  As $\mathcal{S}_f =0$, the isotropic model equations for $k_f$, $k_{f@p}$, $k_p$ and $k_{fp}$ are solved directly:
\begin{subequations}\label{eq:decay}
	\begin{gather}
	\frac{d k_f}{d t} = 
	\frac{2 \varphi}{\tau_p} ( k_{fp} - k_{f@p})
	- \varepsilon_f , \\
	\frac{d k_{f@p} }{d t} = 
	\frac{2 \varphi}{\tau_p} ( k_{fp} - k_{f@p})  
	- \varepsilon_f , \\ 
	\frac{d k_p }{d t}  =
	\frac{2}{\tau_p} ( k_{fp} - k_{p})  
	- \varepsilon_p  , \\
	\frac{d k_{fp} }{d t} =  
	\frac{1}{\tau_p} [ k_{f@p} + \varphi k_p - (1+\varphi) k_{fp} ]
	- \left(
	\frac{1}{2}  + \frac{3}{4} C_{0} \right) \left( \frac{\varepsilon_f}{k_f} + \frac{\varepsilon_p}{k_p} \right) k_{fp} , \\
	\frac{d \varepsilon_f }{d t} = -  C_{\epsilon 2 } \frac{\varepsilon_f^2}{k_f} +  C_{3} \frac{\varphi }{\tau_p} \left( \frac{k_{fp}}{k_{f@p}}  \varepsilon_p - \varepsilon_f \right), \\
	\frac{d \varepsilon_p }{d t}  = - C_{\epsilon 2 }  \frac{\varepsilon_p^2}{k_p} +  \frac{C_{3}}{ \tau_p} \left( \frac{ k_{fp}}{k_{f@p}} \varepsilon_f - \varepsilon_p \right) .
	\end{gather} 
\end{subequations}
In the decaying turbulence test, initial conditions for the simulation are $k_f(0) = k_{f@p}(0) = k_p(0) = k_{fp}(0) = 1.314$, in accordance with the DNS simulation by \cite{sundaram1999numerical}, and $\varepsilon_f(0) = \varepsilon_p(0) = 1.0112$. Moreover, the mass loading is set to $\varphi=0.162$. Note that because $k_f(0) = k_{f@p}(0)$, the first two equations in system \eqref{eq:decay} will yield $k_f(t) = k_{f@p}(t)$ so that only $k_f$ is required to model decaying turbulence for this case. 

\begin{figure}
\centering
\begin{tabular}{cc}
{\includegraphics[height=.35\textwidth,width=.47\textwidth]{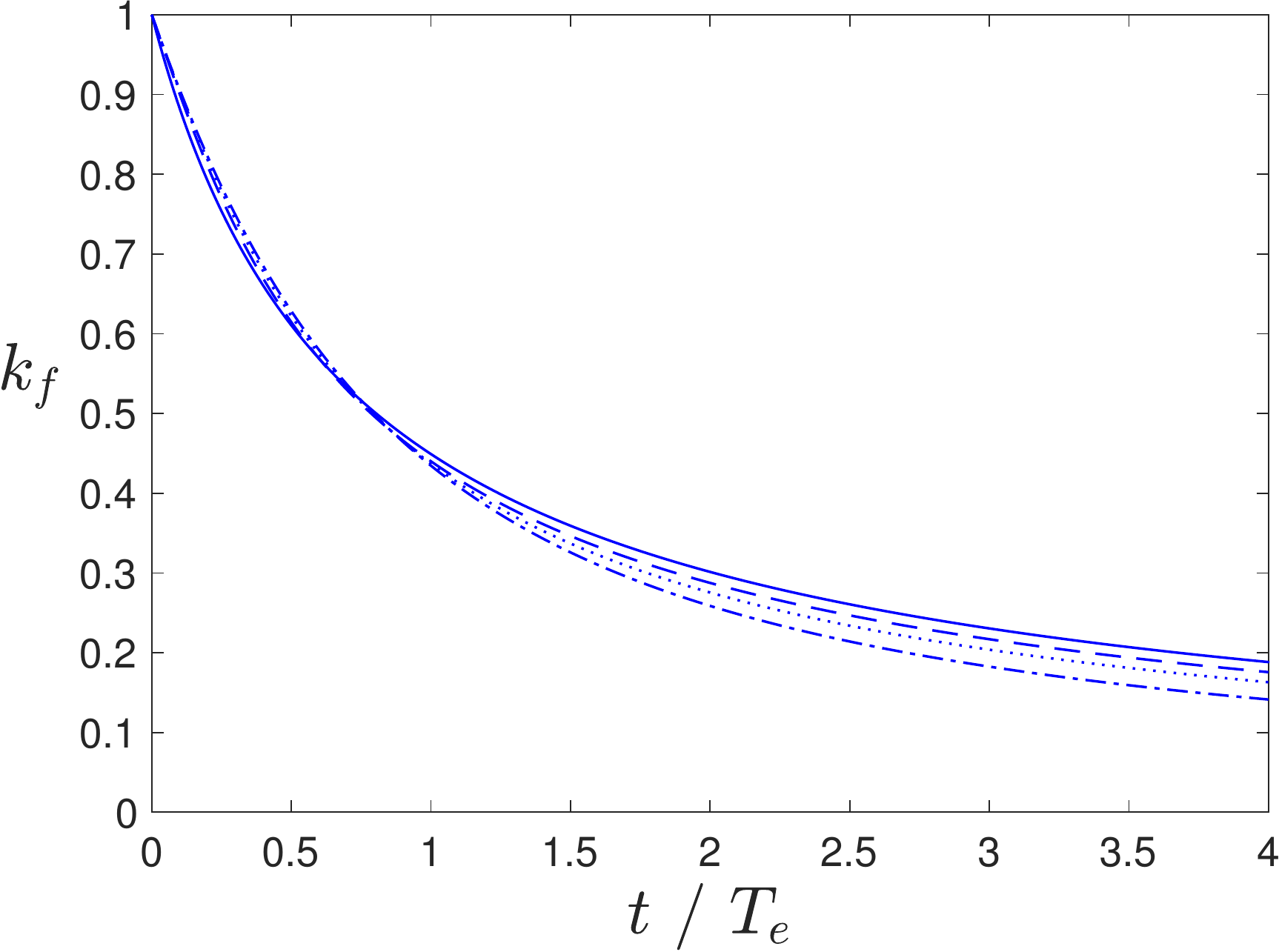}}
\hspace{0.2cm}&
{\includegraphics[height=.35\textwidth,width=.47\textwidth]{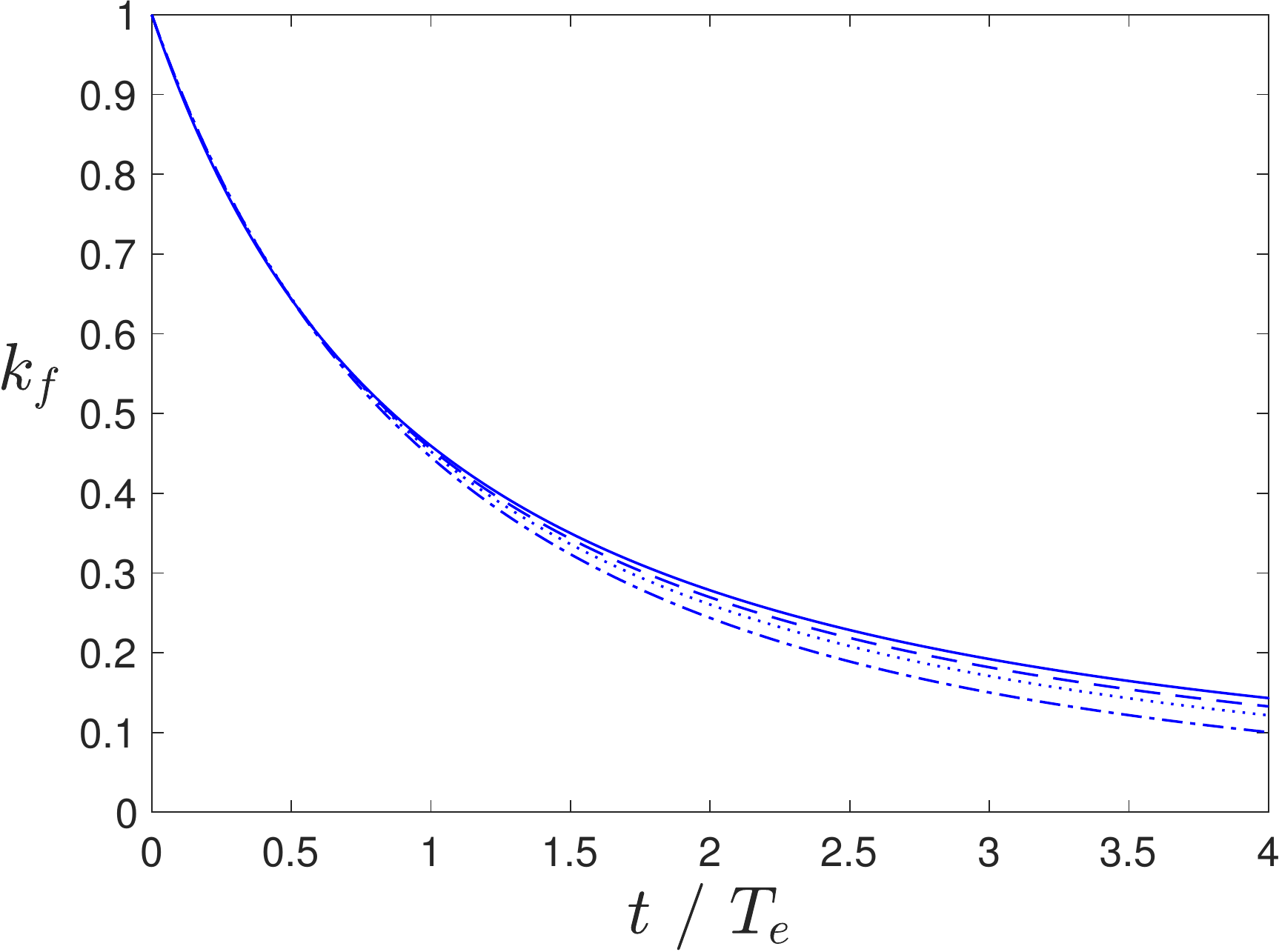}}\\
\hspace{0.5cm}(a) & \hspace{0.5cm}(b)\\ \\
{\includegraphics[height=.35\textwidth,width=.47\textwidth]{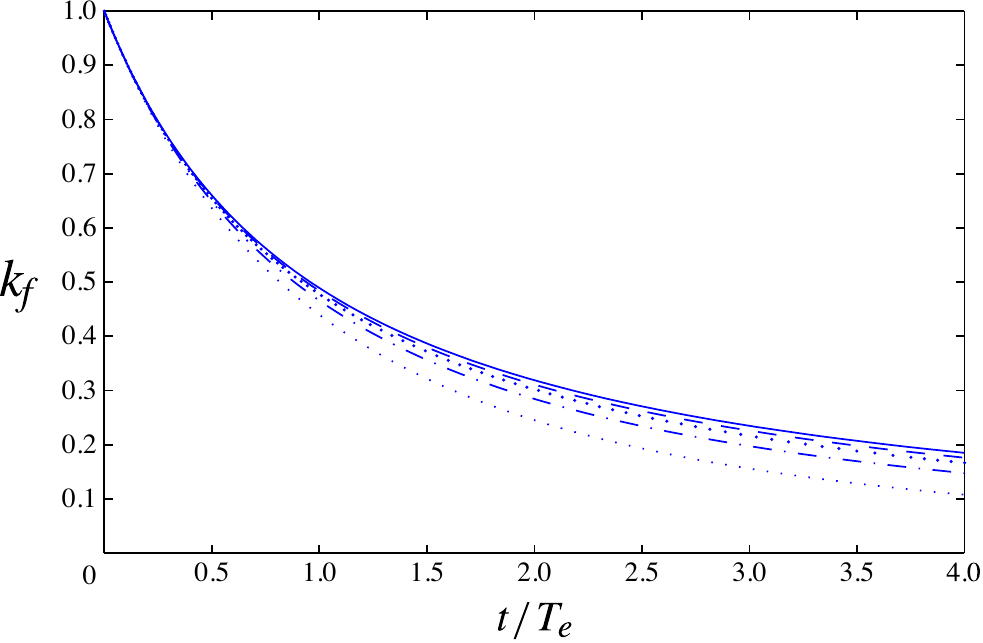}} &
{\includegraphics[height=.36\textwidth,width=.49\textwidth]{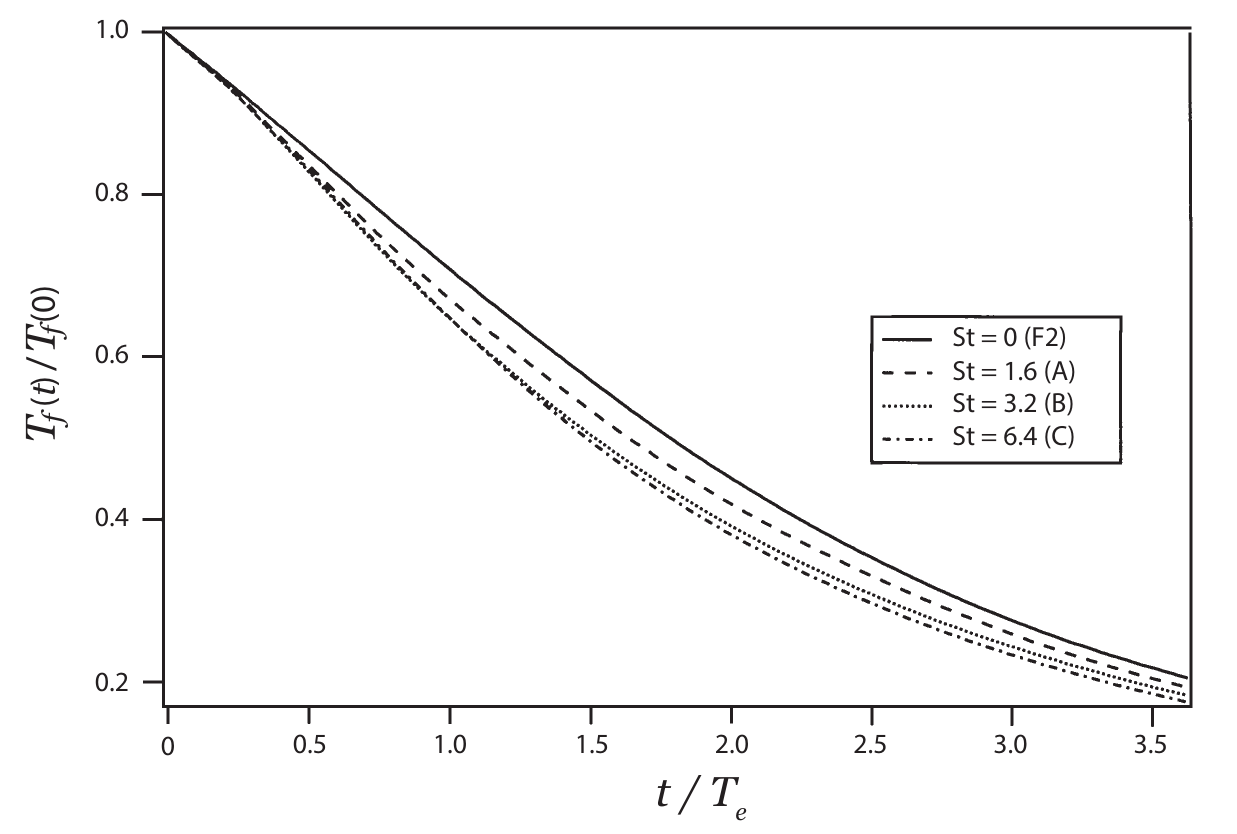}}\\
\hspace{0.5cm}(c) & \hspace{0.5cm}(d)
\end{tabular}
\caption{Fluid turbulent kinetic energy as a function of the non-dimensional time, $t/T_e$, in decaying fluid--particle turbulence: (a) complete particle model, (b) simplified particle model, (c) \cite{Fox2014} Eulerian model and (d) DNS of \cite{sundaram1999numerical}. The curves correspond to four different Stokes number: $St = 0$, solid line; $St = 0.17$, dashed lines; $St= 0.35$, dotted lines; $St= 0.69$, dash-dotted lines.
In panel (c) the light dotted line is relative to an additional value of the Stokes number, not reported in the other panels; in panel (d) $T_f \propto k_f$.}
\label{Fig:decaying}
\end{figure} 
\begin{figure}
\centering
\begin{tabular}{cc}
{\includegraphics[height=.35\textwidth,width=.47\textwidth]{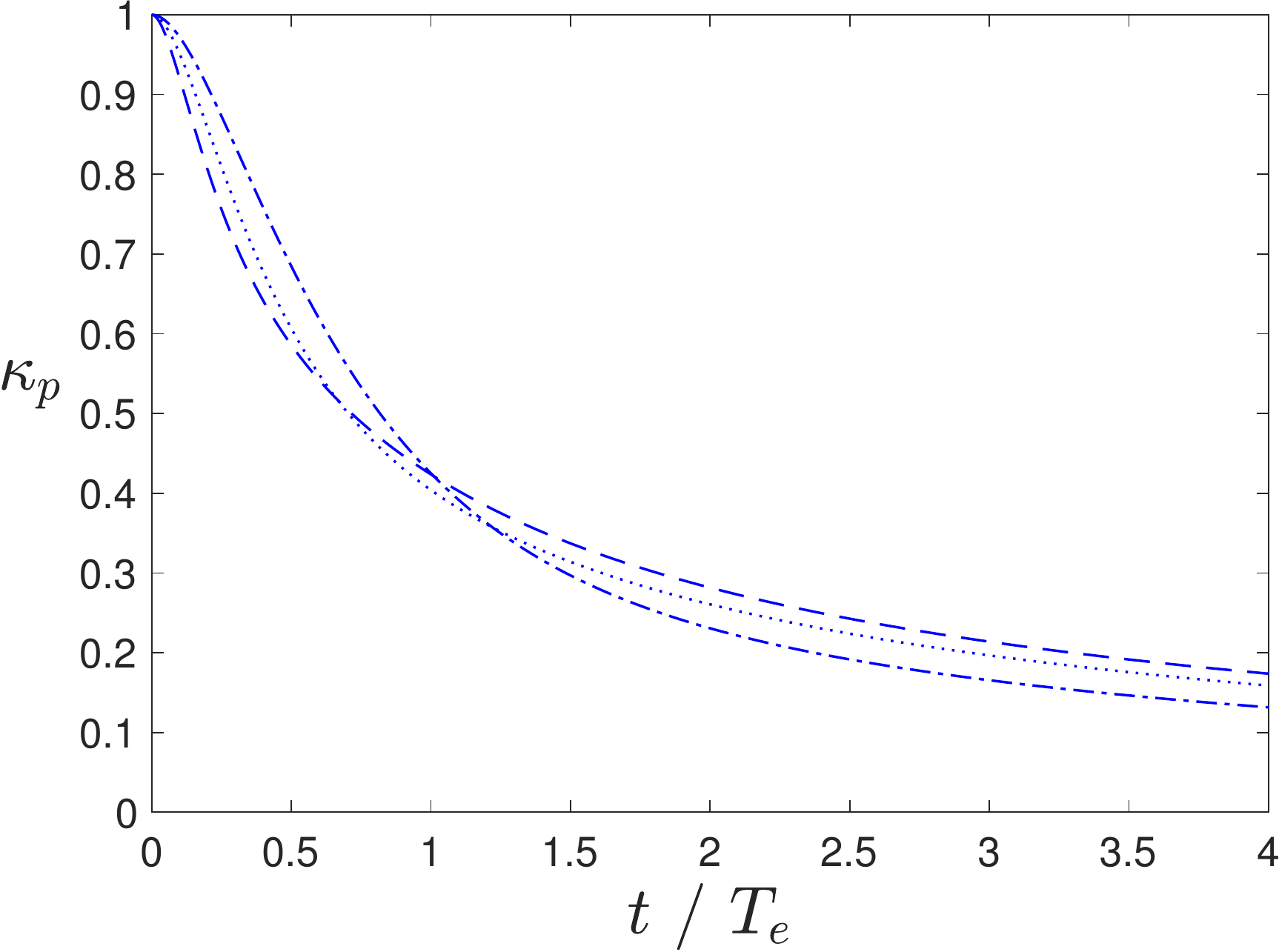}} \hspace{0.2cm}&
{\includegraphics[height=.35\textwidth,width=.47\textwidth]{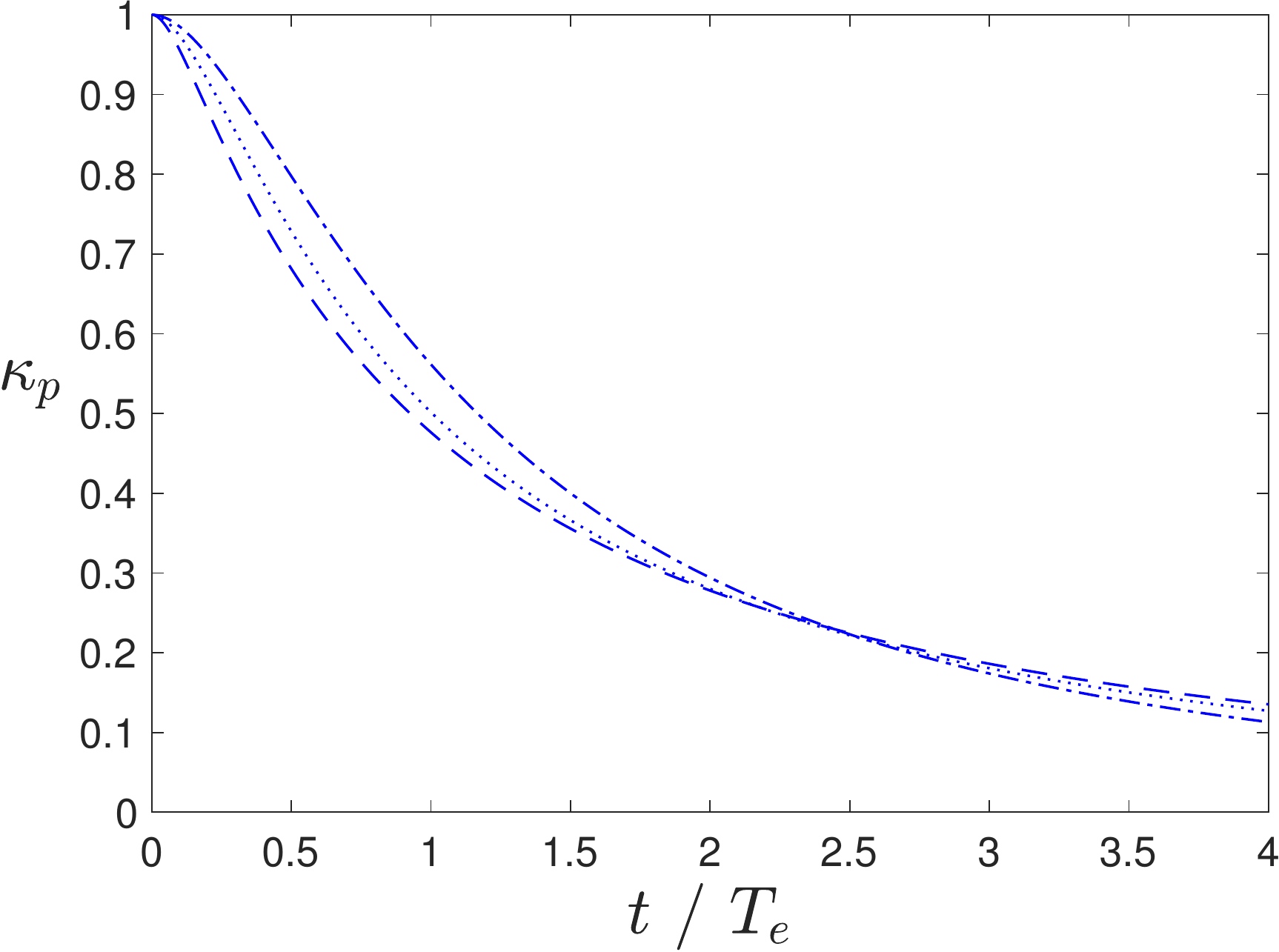}}\\
\hspace{0.5cm}(a) & \hspace{0.5cm}(b)\\ \\
{\includegraphics[height=.35\textwidth,width=.47\textwidth]{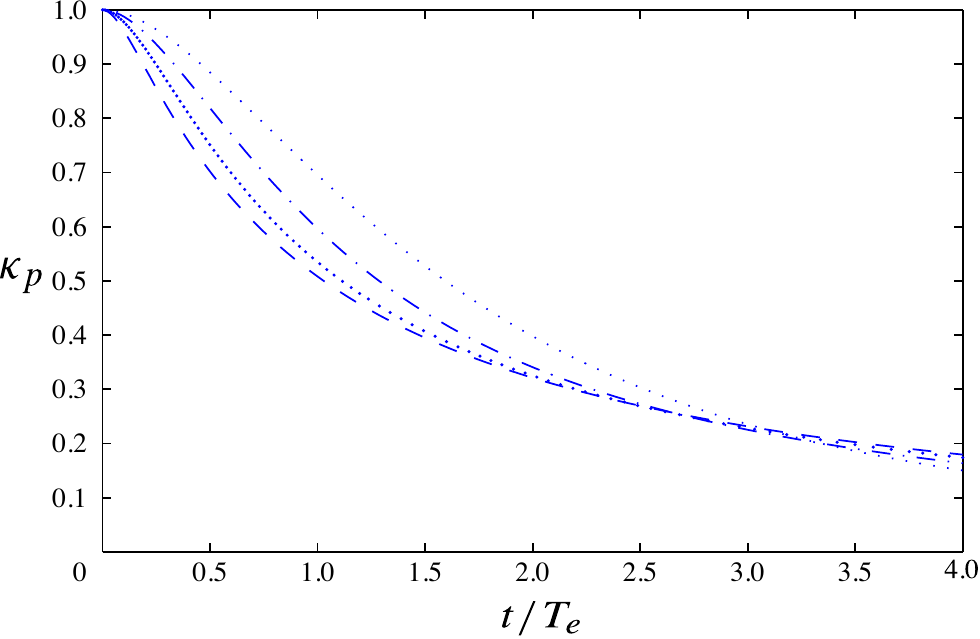}} &
{\includegraphics[height=.36\textwidth,width=.49\textwidth]{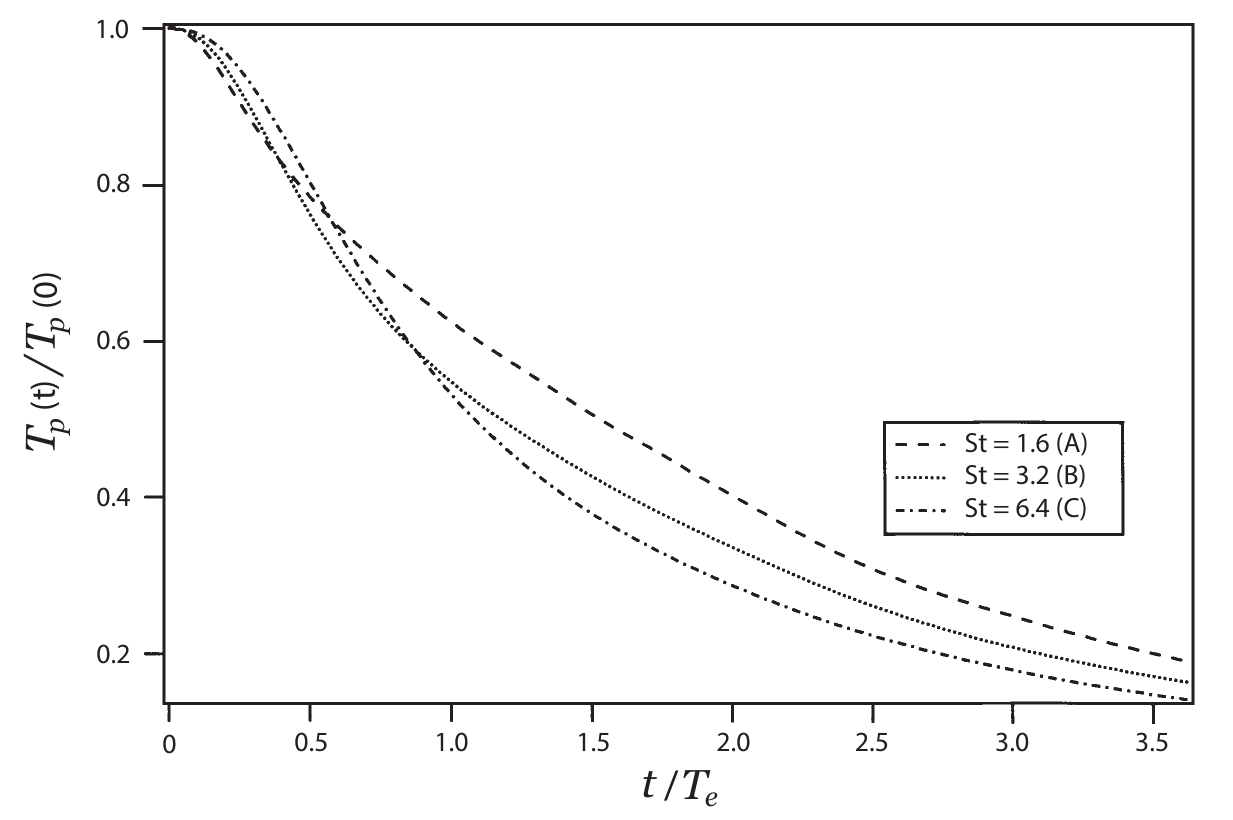}}\\
\hspace{0.5cm}(c) & \hspace{0.5cm}(d)\\ \\
\end{tabular}
\caption{ Particle-phase fluctuating energy  as a function of the non-dimensional time, $t/T_e$, in decaying fluid--particle turbulence: (a) complete particle model, (b) simplified particle model, (c) Eulerian model of \cite{Fox2014} and (d) DNS of \cite{sundaram1999numerical}. The curves correspond to the following Stokes number: $St = 0.17$, dashed lines; $St= 0.35$, dotted lines; $St= 0.69$, dash-dotted lines. In panel (c) the light dotted line is relative to an additional value of the Stokes number, not reported in the other panels; in panel (d) $T_f \propto k_f$.}
\label{Fig:decaying_sund}
\end{figure} 

Figure \ref{Fig:decaying}(a) shows the time evolution of the fluid turbulent kinetic energy obtained with the particle models, for particle sets characterized by four different Stokes numbers, namely $St= \tau_p/ T_e =0$ (fluid tracers), $St=0.17$, $St=0.35$ and $St=0.69$ (where $T_e = 1.7328$ is the initial eddy-turnover time in DNS of \cite{sundaram1999numerical}). Note that the case at $St=0$ was obtained from the particle equations as the limit case for $\tau_p/T_L \ll 1$, as described in Appendix A. Figures \ref{Fig:decaying}(b)--(d) show the same quantities as in figure \ref{Fig:decaying}(a), obtained by the simplified version of the present model, the Eulerian model by \cite{Fox2014} and the DNS by \cite{sundaram1999numerical} respectively. The same comparisons for the particle-phase turbulent kinetic energy are reported in figure \ref{Fig:decaying_sund}. It can be seen that the effect of the Stokes number on the decay of the turbulent kinetic energy of both the fluid and the particle phases is qualitatively well captured by the present stochastic model, in its complete version as well as in the simplified one, although the initial stages of the time evolution are quite different from the DNS results.

\subsubsection{Homogeneous-shear flow}

\begin{figure}
\center
(a){\includegraphics[width=.4\textwidth]{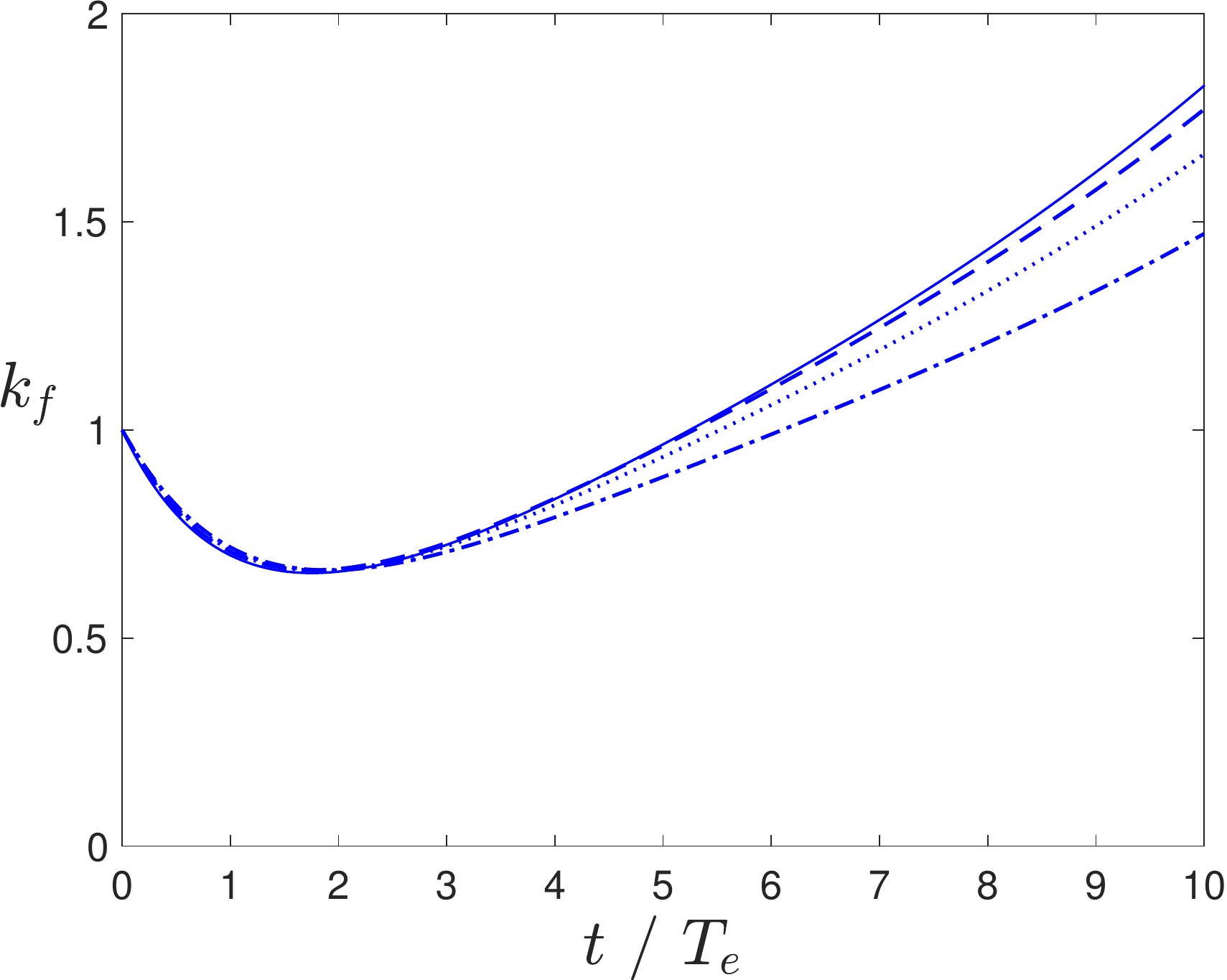}}
\hspace{0.25cm}
(b){\includegraphics[width=.4\textwidth]{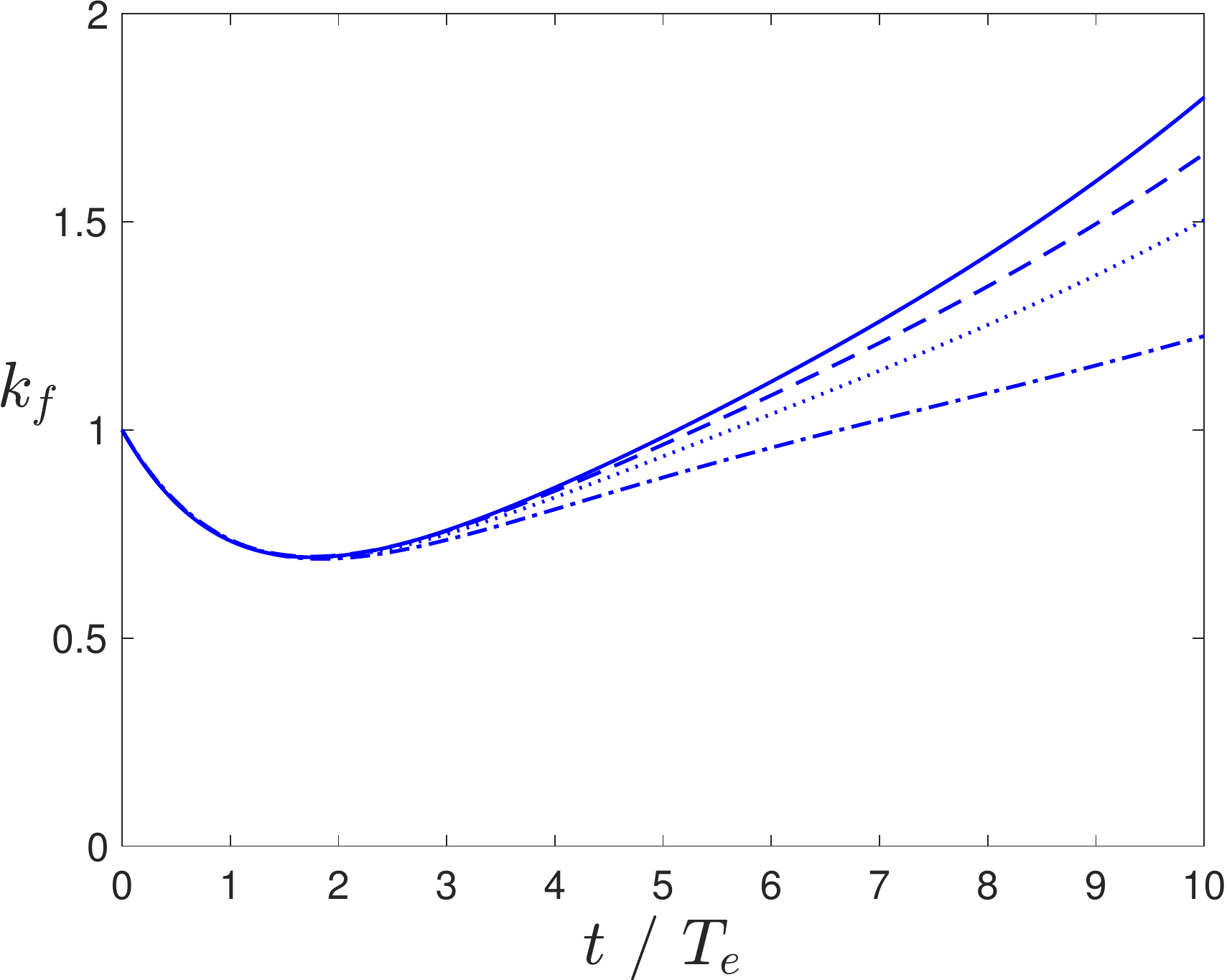}}
\\
(c){\includegraphics[height=.3\textwidth,width=.4\textwidth]{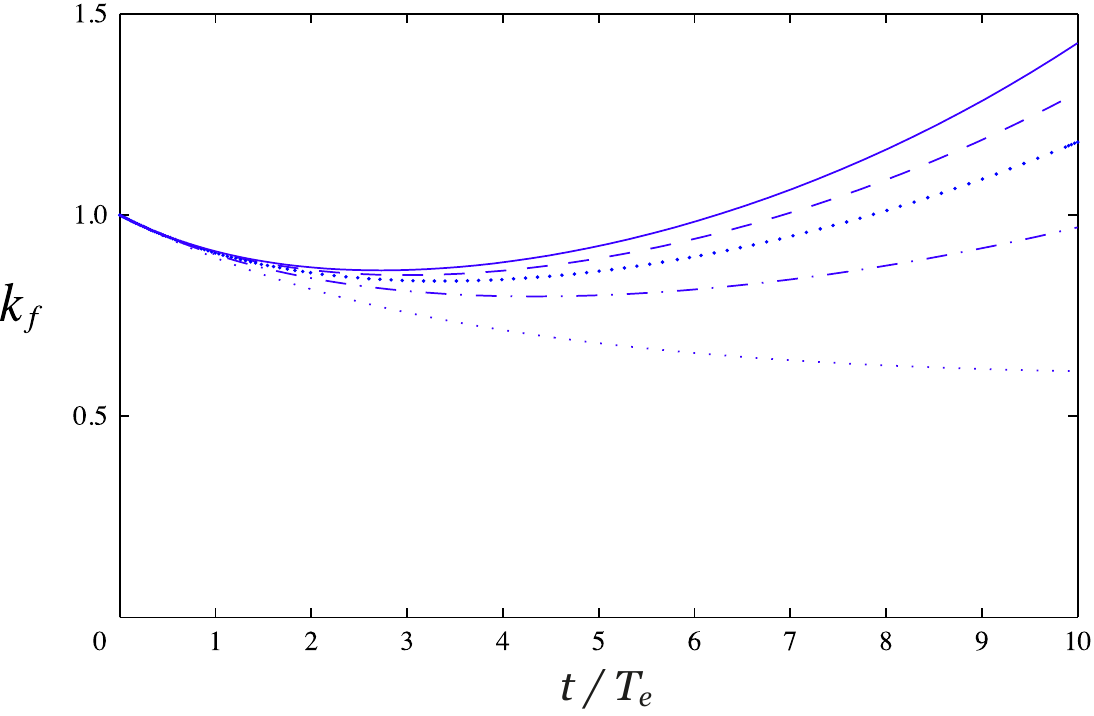}}
\caption{Homogeneous shear flow. Fluid turbulent kinetic energy as a function of the non-dimensional time, $t/T_e$ with (a) the complete stochastic model; (b) the simplified stochastic model and (c) the Eulerian model by \cite{Fox2014}. The curves correspond to four different Stokes numbers: $St = 0$, solid line; $St = 0.17$, dashed lines; $St= 0.35$, dotted lines; $St= 0.69$, dash-dotted lines. In panel (c) the light dotted line is relative to an additional value of the Stokes number, not reported in the other panels.}
\label{Fig:shear}
\end{figure}

\begin{figure}
\centering
(a){\includegraphics[width=.4\textwidth]{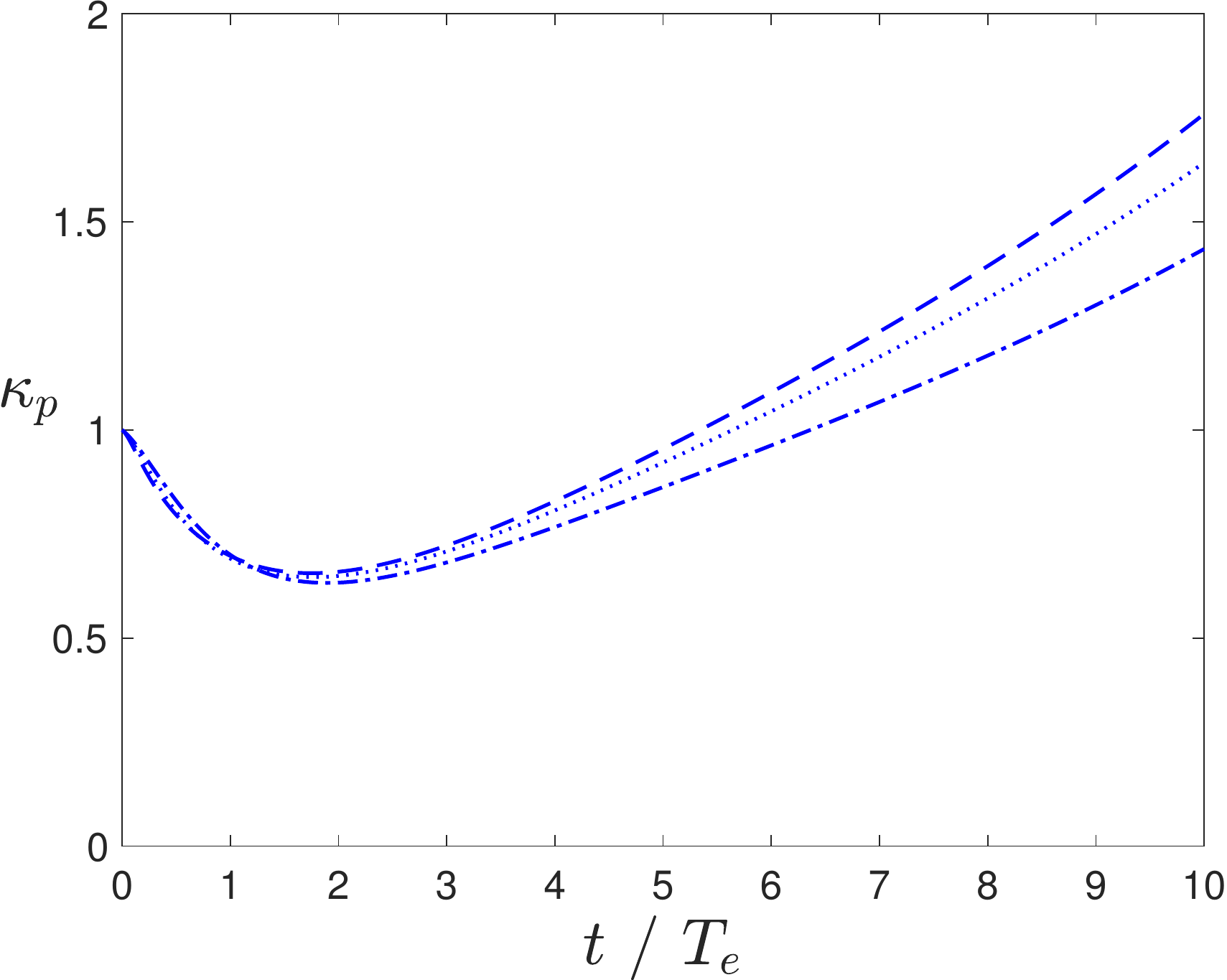}}
\hspace{0.25cm}
(b){\includegraphics[width=.4\textwidth]{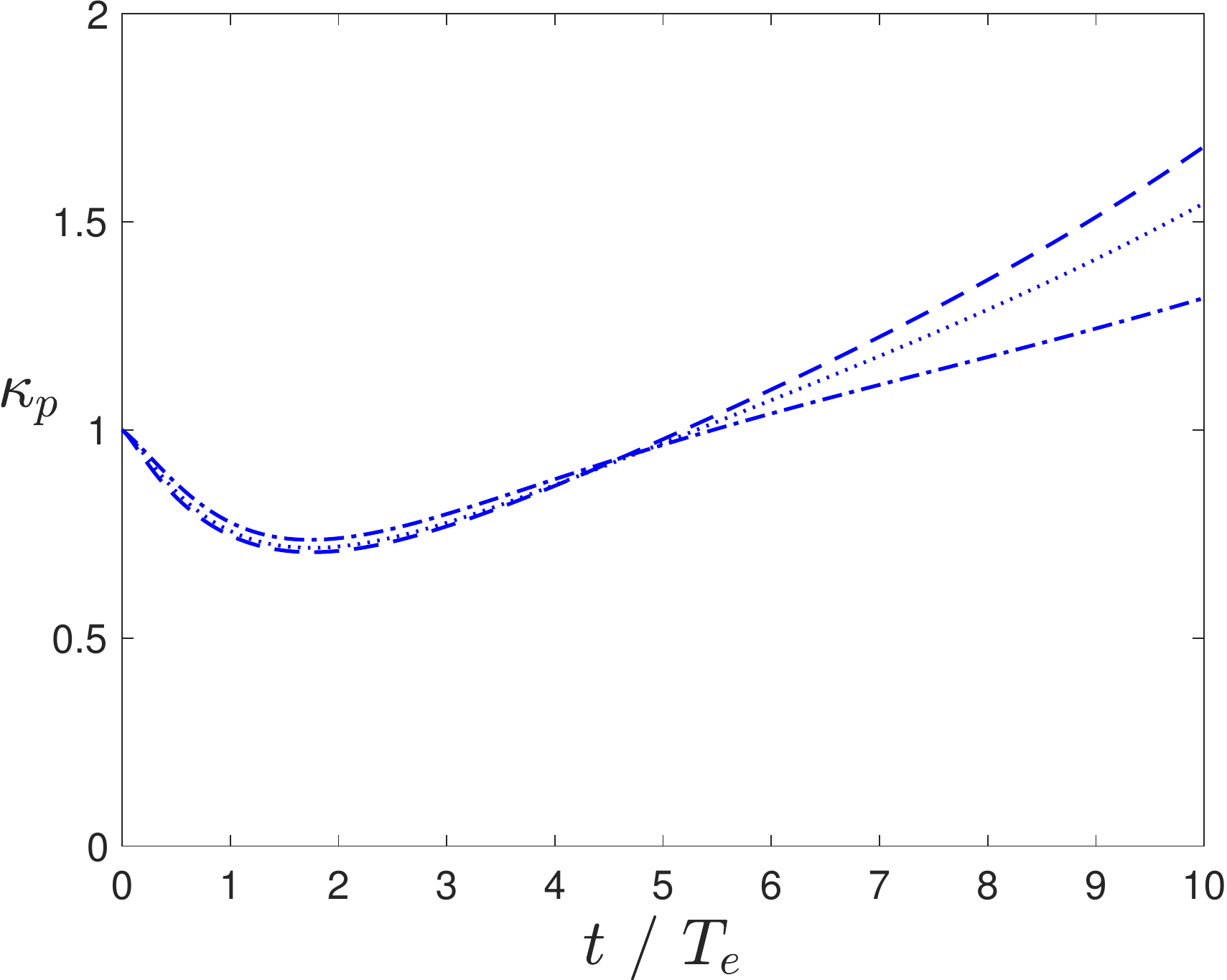}}
\\
(c){\includegraphics[height=.3\textwidth,width=.4\textwidth]{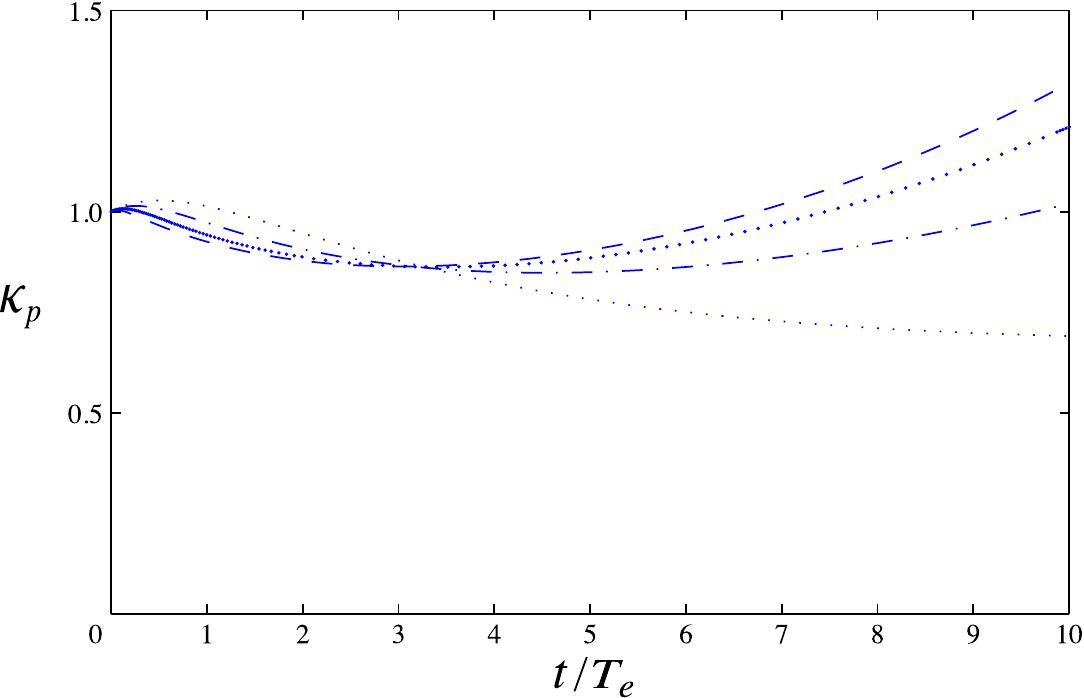}}
%
\caption{Homogeneous shear flow. Particle-phase fluctuating energy as a function of the non-dimensional time, $t/T_e$ with (a) complete particle model; (b) simplified particle model and (c) Eulerian model in \cite{Fox2014}. The curves correspond to the Stokes numbers: $St = 0.17$, dashed lines; $St= 0.35$, dotted lines; $St= 0.69$, dash-dotted lines. In panel (c) the light dotted line is relative to an additional value of the Stokes number, not reported in the other panels.}
\label{Fig:shear_p}
\end{figure}

We consider now the case of a homogeneous shear flow with $\mathcal{S}_f= 0.6$, and solve the anisotropic model equations given in \S\ref{sec:dhsf}. The mass loading is $\varphi = 0.162$ and the initial conditions $\varepsilon_f(0) = \varepsilon_p(0) = 0.25$ (see \cite{Fox2014}). As in the previous decaying case, simulations have been carried out for the following four Stokes numbers: $St = 0, 0.17, 0.35, 0.69$. The values of the constants in our model are the same as in the previous case of homogeneous decaying turbulence with $C_{\epsilon1f} = C_{\epsilon1p} = 1.44$, which are standard values for single-phase turbulence models \citep{Pope_turbulent}. For the simplified model $C_{\epsilon1f} = 1.2$. Figure \ref{Fig:shear} shows the time evolution of the fluid turbulent kinetic energy obtained with the particle model, both in its complete and simplified versions, with the same quantity obtained from the Eulerian model in \cite{Fox2014}. Comparison should also be made with the DNS data in figure 45 of \cite{ahmed2000mechanisms}.  For all the models, the time behaviour is qualitatively similar to that observed in DNS, with an initial decrease of the fluid turbulent kinetic energy followed by an increase. The value of the minima of $k_f$ given by the complete particle model are closer to those obtained in DNS. Moreover, the effect of particle inertia on the time evolution of $k_f$ is also correctly captured, i.e., the rate of increase of the fluid turbulent kinetic energy after the minimum is reduced as the inertia of the particles increases.  The same effect is found also for the particle-phase fluctuating energy, $\kappa_p$, in agreement with \cite{Fox2014}, as it can be seen in Figure \ref{Fig:shear_p}. No DNS data are available for this quantity.


\subsection{Cluster-induced turbulence}

To isolate the effect of turbulence generated by particles through two-way coupling, we consider a flow initially at rest laden with a random distribution of finite-size particles of diameter $d_p$ subject to gravity oriented in the downward $x_1$ direction. The physical parameters are chosen to correspond to the Euler--Lagrange (EL) point-particle simulation of \cite{Fox2015} as summarized in table~\ref{tab:flui-part-param}. The dimensionless two-phase parameters that characterize the flow include the particle-to-fluid density ratio $\rho_p / \rho_f = 1000$, the average particle-phase volume fraction $\langle \alpha_p \rangle = 0.01$ and the particle Reynolds numbers $Re_p = \tau_p g d_p / \nu_f = 1$ where $\tau_p = \rho_p d_p^2/(18\rho_f \nu_f )$ is the particle relaxation time, $\nu_f$ is the fluid-phase kinematic viscosity and $g$ is the magnitude of the gravity vector. Combination of these non-dimensional numbers yields the mass loading $\varphi = \rho_p \langle \alpha_p \rangle/(\rho_f \langle \alpha_f \rangle)=10.1$, where $\langle \alpha_f \rangle=1 - \langle \alpha_p \rangle$ is the average fluid-phase volume fraction. Finally, $\mathcal{V}= g \tau_p$ is the settling velocity for a single particle. 

The CIT case is statistically homogeneous in all directions with periodic boundary conditions; therefore, in the context of the present formalism, it reduces to a 0-D description, with only the time dependency. Moreover, as in the previous considered cases, since the RA equations obtained from the stochastic ones are in closed form for a homogeneous configuration, we can limit ourselves to solving a system of coupled ODEs, instead of carrying out a Lagrangian Monte-Carlo simulation.
The simulation is performed starting from an initial condition where both the particle and fluid phases are at rest and it is evolved in time up to the steady state. The fluid-phase pressure gradient is dynamically adjusted in order to keep the mean fluid velocity $\langle {\bf U}_f \rangle$ equal to zero. 
The model constants have been set in order to obtain a good prediction of the steady-state values for first-order moments. The values, so obtained, are reported in tables~\ref{tab:CIT-constants}--\ref{tab:CIT-constants2}.
{
A comment is in order concerning the values of $C_{0f}$ and $C_{3p}$.
These values are taken different from those used in the isotropic cases previously analysed.
The results obtained in the CIT test-case with the previous values are in reasonable agreement with the full numerical simulation, but show some discrepancy which has been eliminated using the values proposed in table~\ref{tab:CIT-constants}. In fact, $C_{3p}$ has an insignificant effect on the asymptotic results, but the present higher value  smooths the transient dynamics. In contrast, the value of $C_{0f}$ turns out to be key to get the correct level of turbulent kinetic energy.}

\begin{table}
	
\begin{center}
\begin{tabular}{l l l}
\multicolumn{3}{c}{\textbf{Physical parameters}}\\
 $d_p$ & Particle diameter & 0.09 mm  \\
 $\rho_p$ & Particle density & 1000 kg m$^{-3}$ \\
 $\rho_f$ & Fluid density & 1 kg m$^{-3}$ \\
 $\nu_f$ & Fluid kinematic viscosity & $1.8 \bcdot 10^{-5}$ m$^2$ s$^{-1}$ \\
 $g$ & Gravity magnitude & 8 m s$^{-2}$ \\
 \multicolumn{3}{c}{\textbf{Non-dimensional parameters}}\\
 $e$ & Restitution coefficient & 0.90 \\
 $\langle \alpha_p \rangle$ & Mean particle volume fraction & $0.01$ \\
 $\varphi$ & Mean mass loading & $10.1$ \\
 $Re_p$ & Particle Reynolds number & $1$ \\
 \multicolumn{3}{c}{\textbf{Dimensional parameters}}\\
 $\tau_p$ & Drag time & 0.025 s \\
 $\mathcal{V}$ & Settling velocity & 0.20 m s$^{-1} $ \\
\end{tabular}
\end{center}
\caption{Fluid--particle parameters used in CIT simulations \citep{Fox2015}.}
\label{tab:flui-part-param}
\end{table}

\begin{table}
\begin{center}
\begin{tabular}{c c c c c c c c}
$C_{0f}$ & $C_{0p}$ & $C_{\epsilon 2}$ & $C_{3f}$ & $C_{3p}$ & $C_4$ & $f_s$ & $\beta_f$\\
$3.5$ & $0.18$ & $1.92$ & $3.5$ & $7.0$ & $6.81$ & $0.4$ & $1$\\
\end{tabular}
\caption{Values of the model constants used in CIT simulations for the complete model.}
\label{tab:CIT-constants}
\end{center}
\end{table}
\begin{table}
\begin{center}
\begin{tabular}{c c c c c}
$C_{0f}$ & $C_{\epsilon 2}$ & $C_{3f}$ & $C_4$  & $\beta_f$\\
$0.8$ & $6$ & $0.02$ & $0.1$ & $0.75$ \\
\end{tabular}
\caption{Values of the model constants used in CIT simulations for the simplified model.}
\label{tab:CIT-constants2}
\end{center}
\end{table}

The first-order moments are found by solving
\begin{subequations}\label{eq:citmeanU}
\begin{equation}
\frac{d \langle U_{f,1} \rangle}{d t} 
= - \frac{1}{\rho_f \langle \alpha_f \rangle} \frac{d \langle p_f \rangle}{d x_1} + \frac{\varphi}{\tau_p}  \langle U_{p,1} - U_{s,1} \rangle  - g = 0 ,
\label{eq:momentum-homocit}
\end{equation}
\begin{equation}
\label{eq:feq_Up-homcit}
\frac{d \lra{U_{p,1}}}{d t} =
\frac{1}{\tau_p} \langle U_{s,1} - U_{p,1} \rangle  - g,
\end{equation}
\begin{equation}
\label{eq:feq_Us-homcit}
\frac{d \lra{U_{s,1}}}{d t}  =
- \frac{1}{\rho_f}  \frac{d \langle p_f \rangle}{d x_1}  - \frac{1}{T_{L,1}^*}  \langle U_{s,1} \rangle  + \frac{\varphi}{\tau_p}   \langle U_{p,1} - U_{s,1} \rangle - g .
\end{equation}
\end{subequations}
Here, \eqref{eq:momentum-homocit} fixes the fluid pressure gradient in \eqref{eq:feq_Us-homcit}. At steady state, \eqref{eq:feq_Up-homcit} yields $\langle U_{s,1} - U_{p,1} \rangle  = \mathcal{V}$, which agrees with the EL simulations of \cite{Fox2015}, and \eqref{eq:feq_Us-homcit} yields
\begin{equation}
\label{eq:feq_Up-homcitss}
\langle U_{s,1} \rangle  = - \langle \alpha_p \rangle (1 + \varphi) \frac{T_{L,1}^*}{\tau_p} \mathcal{V}
\end{equation}
where, using the definition in \S\ref{sec:part-model}, 
\begin{equation}\label{eq:ssTL}
{ \frac{T_{L,1}^*}{\tau_p} = \left[ \left( \frac{1}{2} + \frac{3}{4} C_{0f} \right)^{2} \Bigl (1 + \frac{3 \beta^2 \mathcal{V}^2}{2 k_f}\Bigl ) \right]^{-1/2} \frac{1}{St_f} }
\end{equation}
and $St_f = \frac{\tau_p \varepsilon_f}{k_f}$. $\beta = T_{Lf}/T_{Ef}$ is set equal to $0.8$.  In fully developed CIT, the turbulence is generated by the clusters and the resulting Stokes number is nearly constant \citep{capecelatro2016}.  The complete model therefore predicts that the steady-state value of $\langle U_{s,1} \rangle /  \mathcal{V}$ depends on the particle volume fraction, the mass loading, and the dimensionless fluid-phase turbulent kinetic energy $2 k_f/ \mathcal{V}^2$. The prediction of $\langle { \bf U}_{s} \rangle$ is perhaps the most important contribution of the Lagrangian pdf model for CIT because information on the fluid seen by the particles is not available in most multiphase turbulence models for fluid--particle flows \citep[see, e.g.,][for details]{Fox2014}.

The second-order moments have two independent, non-zero components, i.e., the vertical $(1,1)$ and horizontal $(2,2)$. For the fluid phase, these are found by solving
\begin{subequations}\label{eq:citRSf}
	\begin{equation}
	\frac{d \langle u_{f,1}^2 \rangle}{d t} 
	= \mathcal{P}_{f,11}  - C_{Rf} \frac{\varepsilon_f}{k_f} \left( \langle u_{f,1}^2 \rangle - \frac{2}{3} k_f \right)  - \frac{2}{3} \varepsilon_f ,
	\label{tkef:cit11}
	\end{equation}
	\begin{equation}
	\frac{d \langle u_{f,2}^2 \rangle}{d t} 
	= \mathcal{P}_{f,22}  - C_{Rf} \frac{\varepsilon_f}{k_f} \left( \langle u_{f,2}^2 \rangle - \frac{2}{3} k_f \right)  - \frac{2}{3} \varepsilon_f 
	\label{tkef:cit22}
	\end{equation}
	where $k_f = \frac{1}{2} ( \langle u_{f,1}^2 \rangle + 2 \langle u_{f,2}^2 \rangle)$, $\mathcal{P}_f = \frac{1}{2} ( \mathcal{P}_{f,11} + 2 \mathcal{P}_{f,22})$,
	\begin{equation}
	\mathcal{P}_{f,11} = \frac{2 \varphi}{\tau_p}  
	[ \langle U_{s,1} \rangle \langle U_{p,1} \rangle {-\langle U_{s,1}\rangle \langle U_{s,1} \rangle}
	+ \langle u_{s,1} u_{p,1} \rangle - \langle u_{s,1}^2 \rangle ] ,
	\end{equation}
	\begin{equation}
	\mathcal{P}_{f,22} = \frac{2 \varphi}{\tau_p} 
     ( \langle u_{s,2} u_{p,2} \rangle - \langle u_{s,2}^2 \rangle )  .
	\end{equation}
\end{subequations}
In CIT,  the fluid-phase Reynolds stresses are anisotropic because of the mean velocities appearing in $\mathcal{P}_{f,11}$.  In general, redistribution is weak so that $\langle u_{f,2}^2 \rangle \ll \langle u_{f,1}^2 \rangle$. 

For the particle-phase pressure tensor, the complete model yields
\begin{subequations}\label{eq:citP}
\begin{equation}
\frac{d \langle P_{11} \rangle}{d t} = \varepsilon_{p,11} - \frac{2 }{\tau_p} \langle P_{11} \rangle  + \frac{1}{2 \tau_c} [ (1+e)^2 \langle \Theta_p \rangle - (1+e)(3-e) \langle P_{11} \rangle ] ,
\end{equation}
\begin{equation}
\frac{d \langle P_{22} \rangle}{d t} = \varepsilon_{p,22} - \frac{2 }{\tau_p} \langle P_{22} \rangle  + \frac{1}{2 \tau_c} [ (1+e)^2 \langle \Theta_p \rangle - (1+e)(3-e) \langle P_{22} \rangle ]  
\end{equation}
where $\langle \Theta_p \rangle = \frac{1}{3} ( \langle P_{11} \rangle + 2 \langle P_{22} \rangle )$,
\begin{equation} 
\varepsilon_{p,11} =  \varepsilon_p \left[ f_s \frac{\langle u_{p,1}^2 \rangle}{k_p} + (1-f_s) \frac{2}{3} \right] ,
\end{equation}
\begin{equation} 
\varepsilon_{p,22} =  \varepsilon_p \left[ f_s \frac{\langle u_{p,2}^2 \rangle}{k_p} + (1-f_s) \frac{2}{3} \right] .
\end{equation}
\end{subequations}
In CIT, the anisotropy of the particle-phase pressure tensor arises due to the source terms $\varepsilon_{p,11}$, $\varepsilon_{p,22}$, whose anisotropy is controlled by $f_s$. For example, if $f_s=1$ and collisions are negligible, the particle-phase pressure tensor and Reynolds stresses will have the same anisotropy.  The collision term, on the other hand, will reduce the anisotropy of the particle-phase pressure tensor.

For the particle phase, the Reynolds stresses are found by solving
\begin{subequations}\label{eq:citRSp}
	\begin{equation}
	\frac{d \langle u_{p,1}^2 \rangle}{d t} 
	= \mathcal{P}_{p,11} - C_{Rp} \frac{\varepsilon_p}{k_p} \left( \langle u_{p,1}^2 \rangle - \frac{2}{3} k_p \right) -  \varepsilon_{p,11} ,
	\label{tkef:cit11p}
	\end{equation}
	\begin{equation}
	\frac{d \langle u_{p,2}^2 \rangle}{d t} 
	= \mathcal{P}_{p,22}  - C_{Rp} \frac{\varepsilon_p}{k_p} \left( \langle u_{p,2}^2 \rangle - \frac{2}{3} k_p \right) -  \varepsilon_{p,22} 
	\label{tkef:cit22p}
	\end{equation}
	where $k_p = \frac{1}{2} ( \langle u_{p,1}^2 \rangle + 2 \langle u_{p,2}^2 \rangle)$, $\mathcal{P}_p = \frac{1}{2} ( \mathcal{P}_{p,11} + 2 \mathcal{P}_{p,22})$,
	\begin{equation}
	\mathcal{P}_{p,11} = \frac{2}{\tau_p}  
	( \langle u_{s,1} u_{p,1} \rangle - \langle u_{p,1}^2 \rangle ) ,
	\label{tkef:cit11ps}
	\end{equation}
	\begin{equation}
	\mathcal{P}_{p,22} = \frac{2}{\tau_p} 
	( \langle u_{s,2} u_{p,2} \rangle - \langle u_{p,2}^2 \rangle )  .
	\label{tkef:cit22ps}
	\end{equation}
\end{subequations}
In CIT, the anisotropy of the particle-phase Reynolds stresses arises due to the production terms, i.e., due to the anisotropy of $\langle u_{s,i} u_{p,j} \rangle$.  The latter are found by solving
\begin{subequations}\label{eq:citRSsp}
	\begin{equation}
	\frac{d \langle u_{s,1} u_{p,1} \rangle}{d t} 
	= \mathcal{P}_{sp,11} - \left( \frac{1}{T_{L,1}^*} + \frac{1}{T_{Lp,1}}  \right) \langle u_{s,1} u_{p,1} \rangle  ,
	\label{tkef:cit11sp}
	\end{equation}
	\begin{equation}
	\frac{d \langle u_{s,2} u_{p,2} \rangle}{d t} 
	= \mathcal{P}_{sp,22} - \left( \frac{1}{T_{L,2}^*} + \frac{1}{T_{Lp,2}}  \right) \langle u_{s,2} u_{p,2} \rangle 
	\label{tkef:cit22sp}
	\end{equation}
	where $k_{fp} = \frac{1}{2} ( \langle u_{s,1} u_{p,1} \rangle + 2 \langle u_{s,2} u_{p,2} \rangle)$, $\mathcal{P}_{sp} = \frac{1}{2} ( \mathcal{P}_{sp,11} + 2 \mathcal{P}_{sp,22})$,
	\begin{equation}
	\mathcal{P}_{sp,11} = \frac{1}{\tau_p} 
	\left[ \lra{ {u}_{s,1}^2 } - \lra{ {u}_{s,1} {u}_{p,1} } + \varphi ( \lra{{u}_{p,1}^2 } - \lra{ {u}_{s,1} {u}_{p,1} } ) \right] ,
	\end{equation}
	\begin{equation}
	\mathcal{P}_{sp,22} = \frac{1}{\tau_p} 
	\left[ \lra{ {u}_{s,2}^2 } - \lra{ {u}_{s,2} {u}_{p,2} } + \varphi ( \lra{{u}_{p,2}^2 } - \lra{ {u}_{s,2} {u}_{p,2} } ) \right]  .
	\end{equation}
\end{subequations}

Likewise, the Reynolds stresses for the fluid seen by the particles are found from
\begin{subequations}\label{eq:citRSs}
	\begin{equation}
	\frac{d \lra{{u}_{s,1}^2} }{d t} = \mathcal{P}_{f,11}   
	+ \frac{2}{{ T}_{L,1}^*} \left( \langle { U}_{s,1} \rangle^2 +  \frac{2}{3} \tilde{k}_f - \lra{ {u}_{s,1}^2 } \right) 
	- \frac{2}{3} \varepsilon_f,
	\label{tkef:cit11s}
	\end{equation}
	\begin{equation}
	\frac{d \lra{{u}_{s,2}^2} }{d t} = \mathcal{P}_{f,22}  
	+ \frac{2}{T_{L,2}^*} \left( \frac{2}{3} \tilde{k}_f - \lra{ {u}_{s,2}^2 } \right) 
	- \frac{2}{3} \varepsilon_f
	\label{tkef:cit22s}
	\end{equation}
	where $\tilde{k}_f$ and $T_{L,i}^*$ are defined in \S\ref{sec:part-model}.  
\end{subequations}
As seen for the fluid phase, the anisotropy of fluid seen is mainly due to the production terms, but is also due to the directional dependence of $T_{L,i}^*$. The dissipation rates $\varepsilon_p$ and $\varepsilon_f$ are found by solving \eqref{eq:epsilon_p1} and \eqref{eq:epsilon2-homo} with the mean-shear-production terms set to zero.

\begin{figure}
\center
\begin{tabular}{cc}
{\includegraphics[width=.38\textwidth]{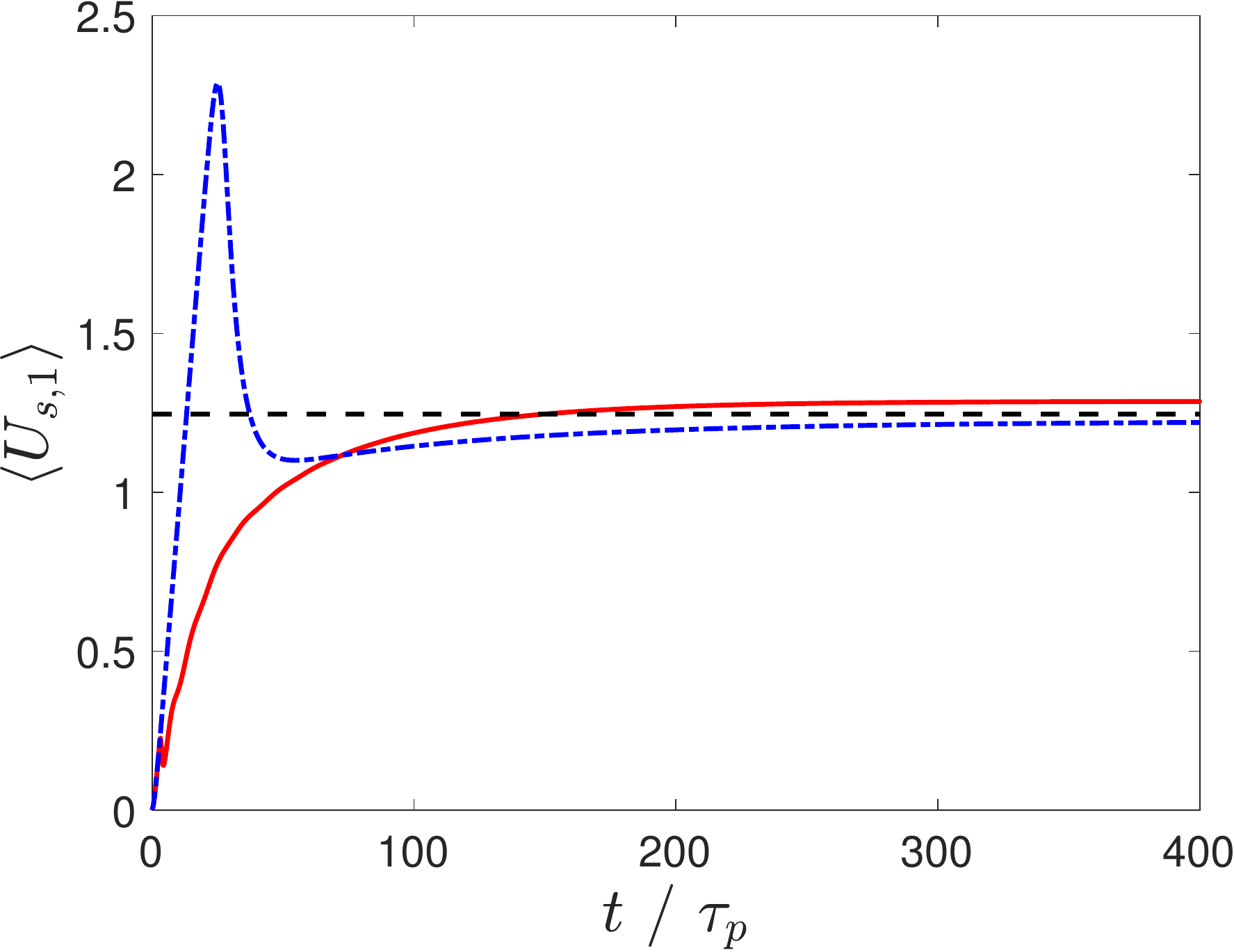}}
\hspace{1cm} &
{\includegraphics[width=.38\textwidth]{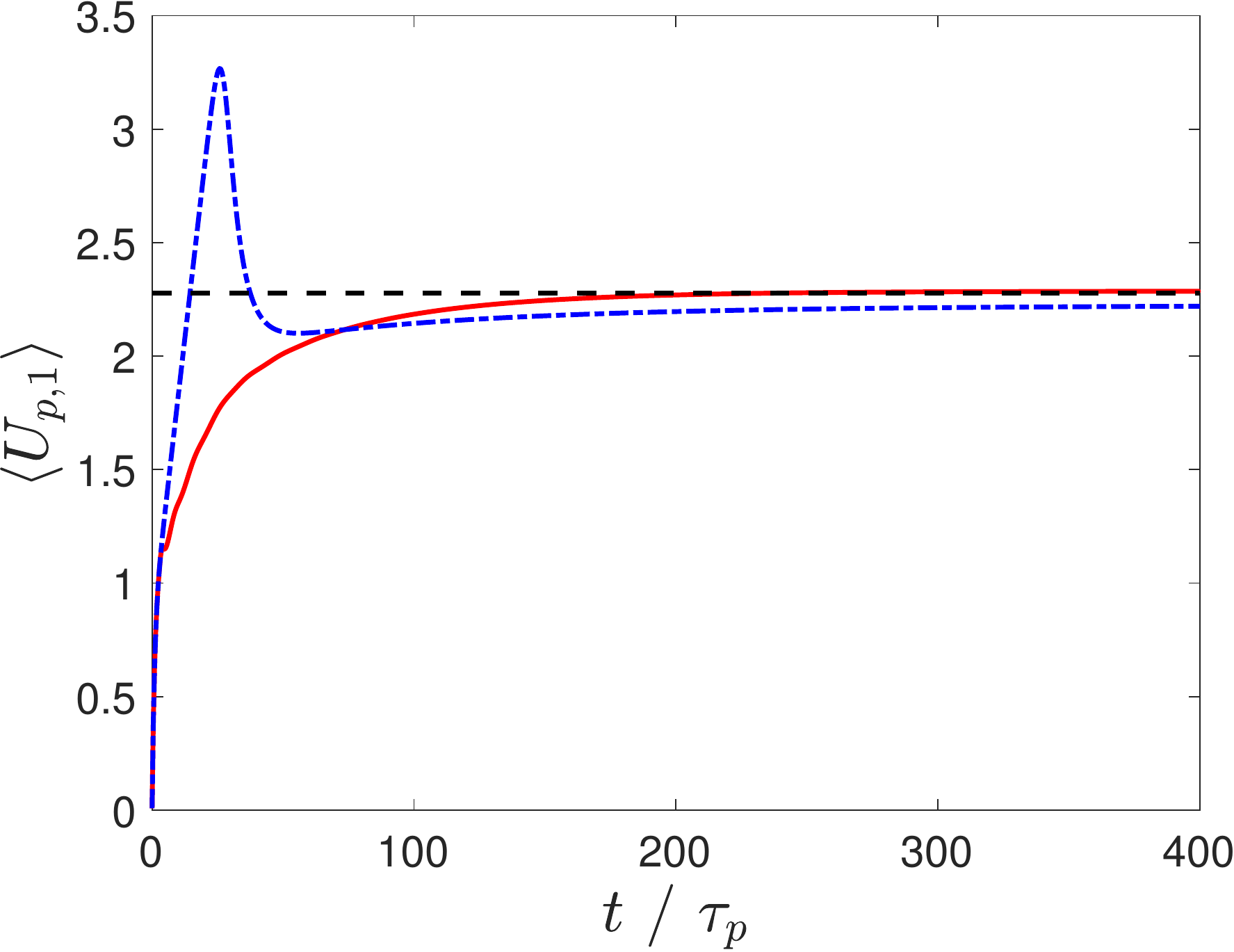}}\\
(a) & (b)
\end{tabular}
\caption{Time evolution of the vertical mean fluid velocity seen by the particles (a) and of the vertical mean particle velocity (b)  from the complete (red line) and from the simplified (blue dot-dashed line) stochastic model.}
\label{Fig:mean_time_diss}
\end{figure}

\begin{figure}
\center
\begin{tabular}{cc}
{\includegraphics[width=.38\textwidth]{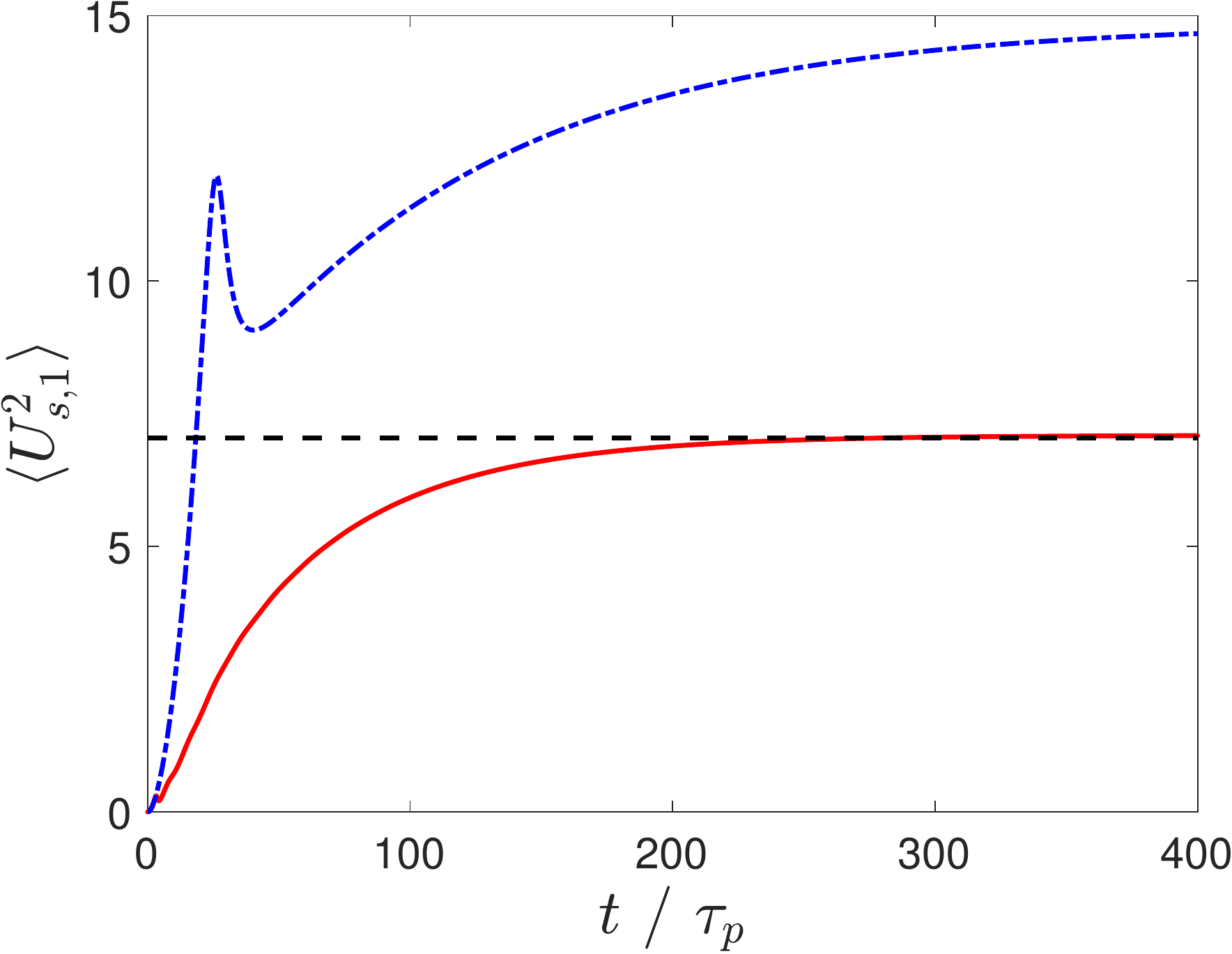}}
\hspace{1cm} &
{\includegraphics[width=.38\textwidth]{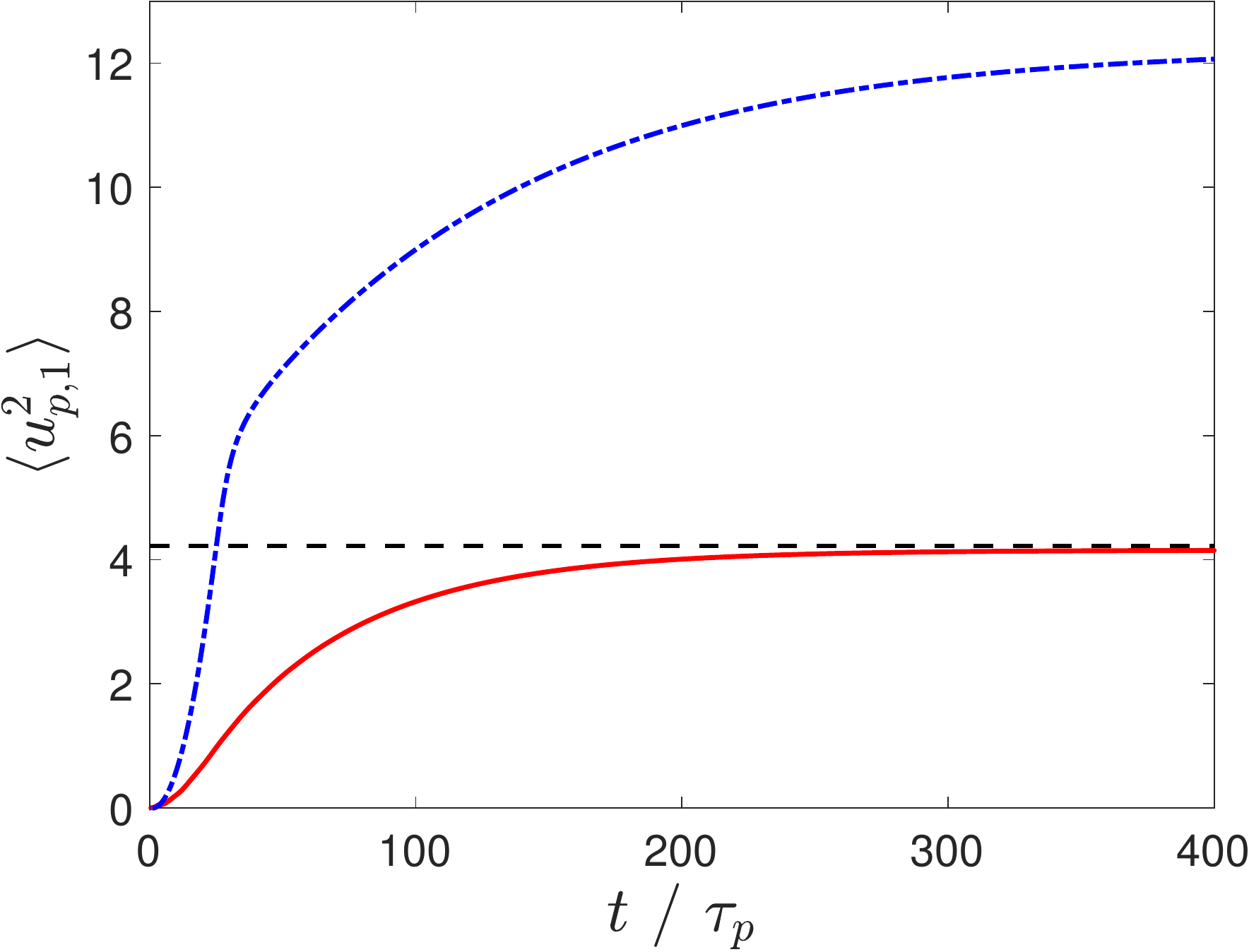}}\\
(a) & (b)
\end{tabular}
\caption{Time evolution of second-order moments of the vertical mean fluid velocity seen by the particles (a) and of the vertical mean particle velocity (b)  from the complete (red line) and from the simplified (blue dot-dashed line) stochastic model.}
\label{Fig:sqr_time_diss}
\end{figure}

\begin{figure}
\center
\begin{tabular}{cc}
{\includegraphics[width=.38\textwidth]{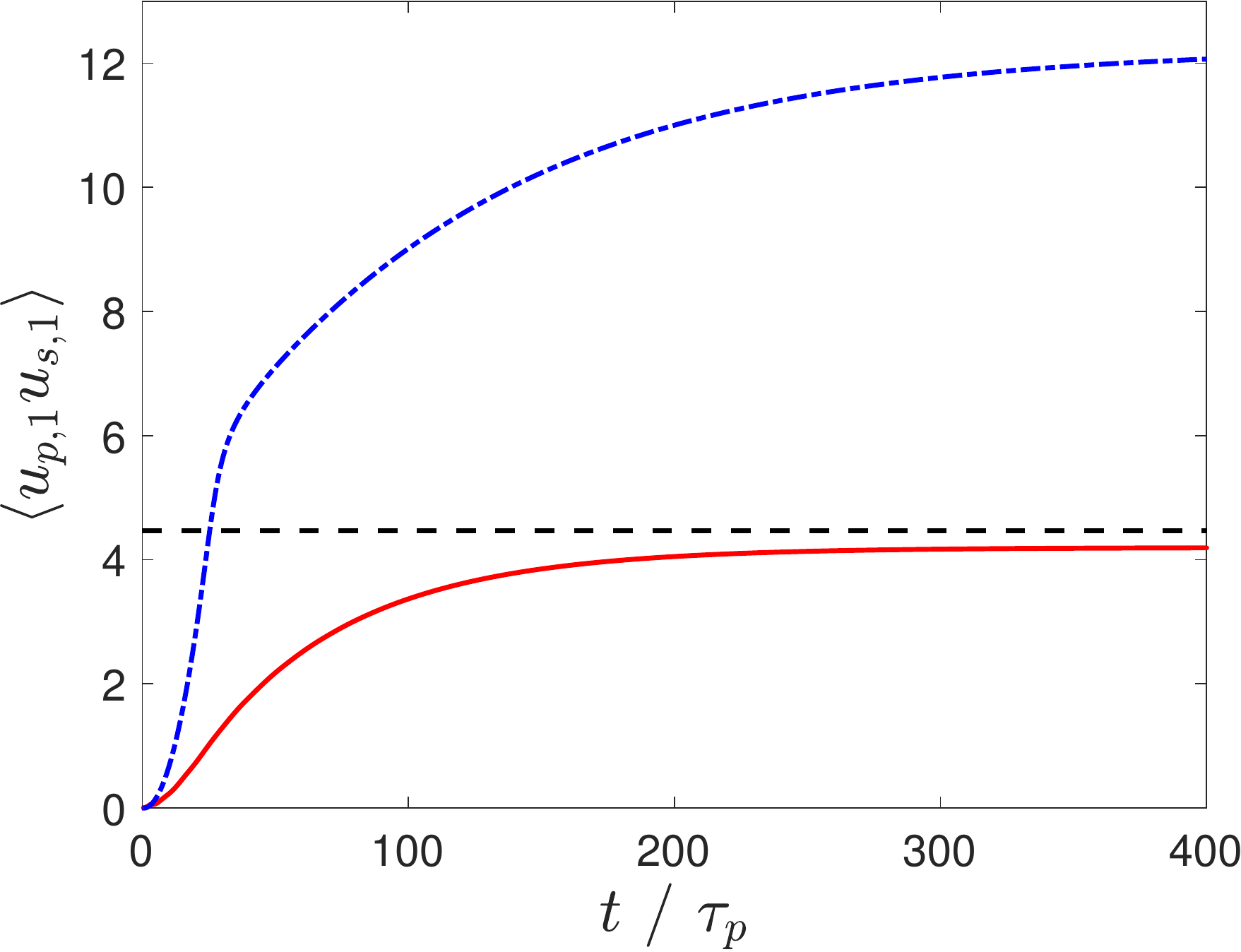}}
\hspace{1cm} &
{\includegraphics[width=.38\textwidth]{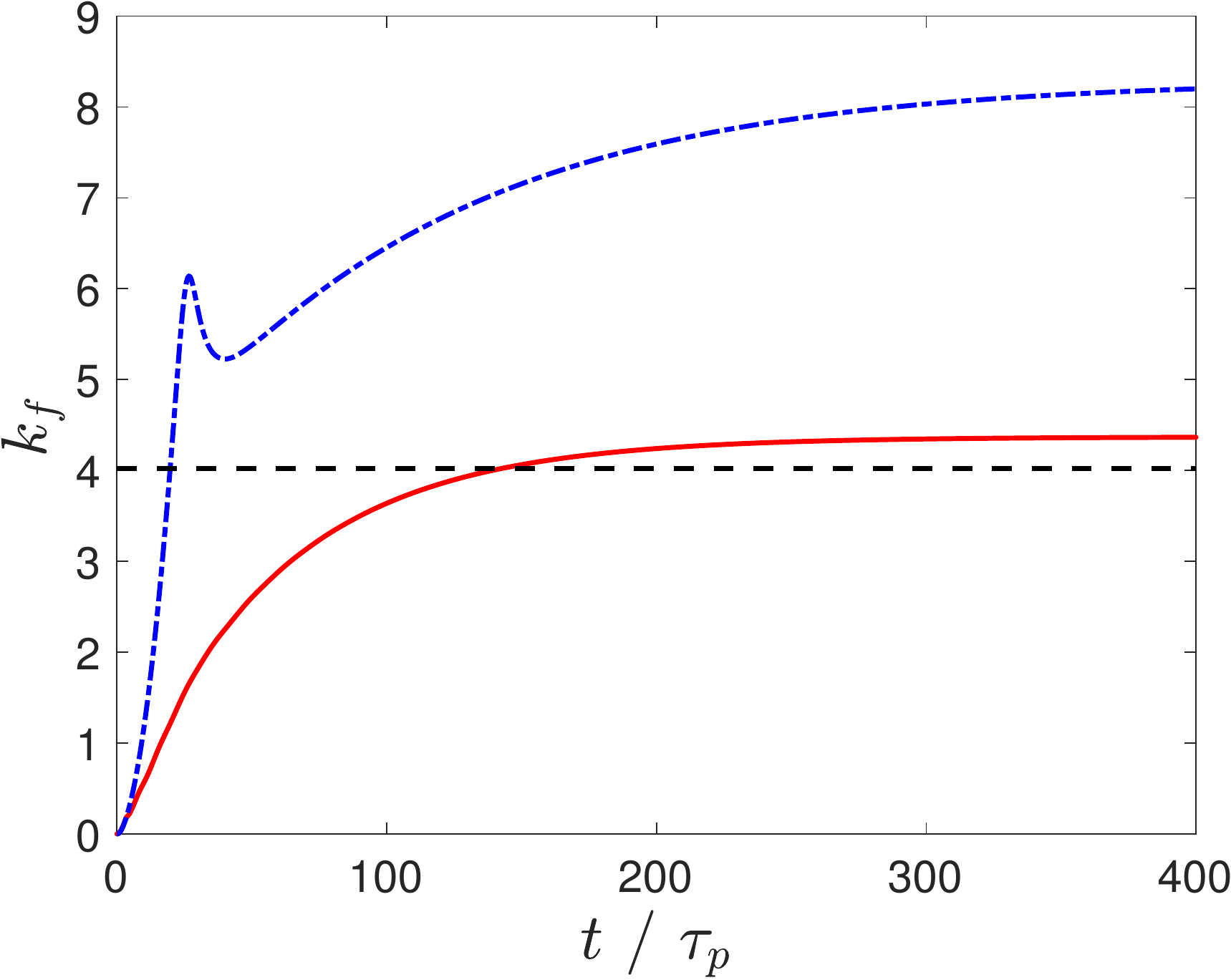}}\\
(a) & (b)
\end{tabular}
\caption{Time evolution of $\langle u_{p,1} u_{s,1} \rangle$ (a) and of $k_f$ (b)  from the complete (red line) and from the simplified (blue dot-dashed line) stochastic model. }
\label{Fig:turb_time_diss}
\end{figure}

Figures \ref{Fig:mean_time_diss}--\ref{Fig:turb_time_diss} show the time evolution of some mean velocities and second-order moments of both the fluid and the particle phase, obtained with the complete and simplified models. It can be seen that all the quantities, after a transient of about 80--100$\tau_p$ due to the non-trivial coupling between particles and fluid, tend to a steady value. The dashed horizontal line in the figures is the steady-state value obtained in the EL simulation by \cite{Fox2015}. It can be seen that the mean velocities (figure \ref{Fig:mean_time_diss}) are well captured by both the complete and simplified models. In particular, an important feature of the Lagrangian models is to provide a prediction of the fluid velocity seen by the particles, as compared to Eulerian models in which it must be a-priori specified. As for the second-order moments (figures \ref{Fig:sqr_time_diss} and \ref{Fig:turb_time_diss}), the complete model still gives a good agreement with EL simulations for both the fluid and particle phases, while the simplified version significantly overestimates the steady-state values. 

Tables \ref{tab:fluid-statistics}--\ref{tab:fluidseen-statistics}, in which the steady-state values of particle and fluid statistics are reported, confirm the previous observations. Table \ref{tab:particle-statistics} also shows the repartition of the particle turbulent kinetic energy, $\kappa_p$, into the coherent part, $k_p$, and granular temperature, $\Theta_p$. For the simplified model, by definition, $\Theta_p=0$ and $\kappa_p=k_p$. The complete model also underestimates the granular temperature, most likely due to underestimating the value of $\varepsilon_p$ through the choice of $C_{3p}$. Nonetheless, the decomposition of the particle turbulent kinetic energy appears to be essential to well predict second-order statistics, as done by the complete model in contrast to the simplified one. The complete model also well reproduces the fact that for all the quantities, except for the uncorrelated part of the particle velocity, turbulence fluctuations in the vertical direction are much higher than those in the horizontal directions. The complete model is able to correctly reproduce the anisotropy of the second-order tensors while the simplified model only gives a qualitative agreement.

\begin{table}
\begin{center}
\begin{tabular}{c c c c}
& $2 k_f / \mathcal{V}^2$ & $\langle u_{f,1}^2 \rangle / (2 k_f)$ & $\langle u_{f,2}^2 \rangle / (2 k_f)$  \\
EL simulation & $8.04$ & $0.82$ & $0.09$  \\
Complete model & $8.74$ & $0.93$ & $0.04$  \\
Simplified model & $16.60$ & $0.98$ & $0.01$  \\
\end{tabular}\\
\caption{Steady-state values of fluid-phase velocity statistics. EL simulation data are taken from \cite{Fox2015}. }
\label{tab:fluid-statistics}
\end{center}
\end{table}

\begin{table}
\begin{center}
\begin{tabular}{c c c c}
& $\langle U_{p,1} \rangle / \mathcal{V}$ & &\\
EL simulation &  $-2.28$ & &\\
Complete model & $-2.28$ & &\\
Simplified model & $-2.22$ & &\\ \\
& $2 \kappa_p / \mathcal{V}^2$ & $\langle v_{p,1}^2 \rangle / (2 \kappa_p)$ & $\langle v_{p,2}^2 \rangle / ( 2 \kappa_p)$  \\
EL simulation & $5.41$ & $0.78$ & $0.11$  \\
Complete model & $5.13$ & $0.81$ & $0.09$  \\
Simplified model & $12.71$ & $0.96$ & $0.02$  \\ \\
& $ k_p / \kappa_p$ & $\langle u_{p,1}^2 \rangle / (2 k_p)$ & $\langle u_{p,2}^2 \rangle / ( 2 k_p)$  \\
EL simulation & $0.89$ & $0.81$ & $0.09$  \\
Complete model & $0.99$ & $0.81$ & $0.09$  \\\\
& $3 \langle \Theta_p \rangle / ( 2 \kappa_p) $ & $\langle P_{11} \rangle / (3 \langle \Theta_p \rangle)$ & $\langle P_{22} \rangle / ( 3 \langle \Theta_p \rangle)$  \\
EL simulation & $0.11$ & $0.51$ & $0.25$  \\
Complete model & $0.01$ & $0.49$ & $0.25$  \\
\end{tabular}\\
\caption{CIT - Steady-state values of particle-phase velocity statistics. EL simulation data are taken from \cite{Fox2015}. }
\label{tab:particle-statistics}
\end{center}
\end{table}

\begin{table}
\begin{center}
\begin{tabular}{c c c c}
& $\langle U_{s,1} \rangle  / \mathcal{V} $ & &\\
EL simulation &  $-1.25$ & &\\
Complete model & $-1.28$ & &\\
Simplified model & $-1.22$ & &\\ \\
& $2 k_{f@p} / \mathcal{V}^2$ & $\langle u_{s,1}^2 \rangle / (2 k_{f@p})$ & $\langle u_{s,2}^2 \rangle / ( 2 k_{f@p})$  \\
EL simulation & $8.32$ & $0.85$ & $0.07$  \\
Complete model & $8.06$ & $0.88$ & $0.06$  \\
Simplified model & $15.36$ & $0.96$ & $0.02$  \\ \\
& $ 2 k_{fp} / \mathcal{V}^2$ & $\langle u_{s,1} u_{p,1} \rangle / (2 k_{fp})$ & $\langle u_{s,2} u_{p,2} \rangle / ( 2 k_{fp})$  \\
EL simulation & $5.45$ & $0.82$ & $0.09$  \\
Complete model & $5.13$ & $0.82$ & $0.09$  \\
Simplified model & $12.71$ & $0.96$ & $0.02$ 
\end{tabular}\\
\caption{CIT - Steady-state values of fluid-phase turbulence statistics seen by particles. EL simulation data are taken from \cite{Fox2015}.}
\label{tab:fluidseen-statistics}
\end{center}
\end{table}

\begin{figure}
\center
{\includegraphics[width=.38\textwidth]{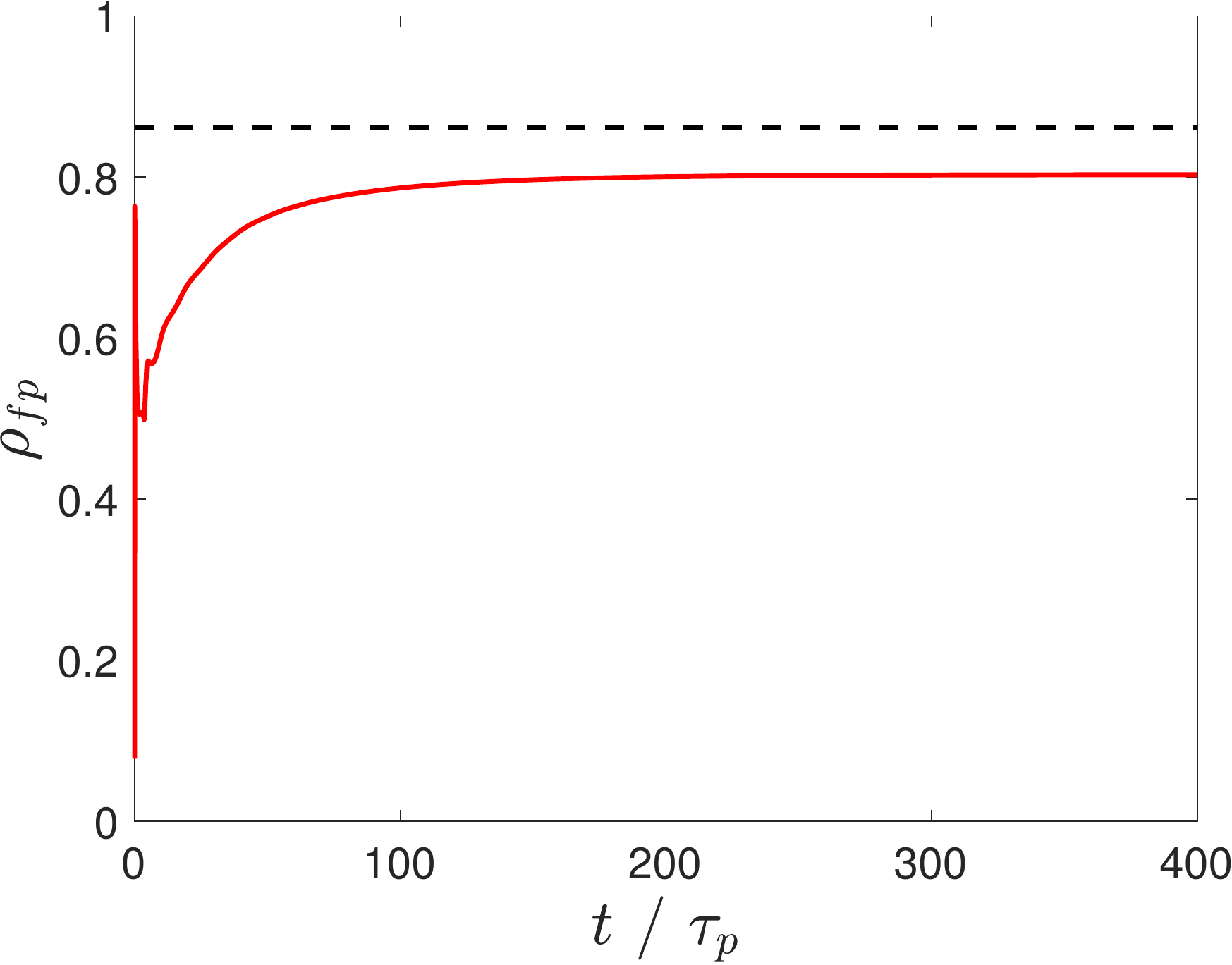}}
\caption{Time evolution of $\rho_{fp} = k_{fp} / (k_{f@p} k_p)^{1/2}$ from the complete stochastic model. }
\label{Fig:correlation}
\end{figure}

Finally in figure \ref{Fig:correlation} it is shown the time evolution of the correlation coefficient 
\begin{equation}
\rho_{fp} = \frac{k_{fp} }{ \sqrt{k_{f@p} k_p}}
\end{equation}
which proofs the importance of having a stochastic model that predicts $k_{fp}$, leading to a correlation coefficient $\rho_{fp}$ that can vary in time, instead of setting it to a constant value.

\section{Conclusions and discussion}\label{sec::Con}

In this work, the stochastic models developed in Part~I have been applied to statistically homogeneous particle-laden flows of increasing complexity. Compared to previous models for strongly coupled flows where such statistics are approximated \citep{Fox2014,capecelatro2016}, the models used here explicitly account for the fluid statistics seen by the particles. While this approach introduces more variables, namely $\mathbf{U}_s$, it eliminates the need to close the coupling terms between the particle and fluid phases. For homogeneous flows with one-way coupling, the differences between the stochastic models and previous models is small.  However, for the cases with two-way coupling, and especially with non-zero mean-slip velocity as in CIT, the correct prediction of $\mathbf{U}_s$ is crucial for successful overall predictions. Interestingly, the steady-state model for $\langle \mathbf{U}_s \rangle$ given in \eqref{eq:feq_Up-homcitss} is relatively simple (compare, for example, the correlation used in \citep{capecelatro2016}), with the Lagrangian time scale $T^*_{L,1}$ playing a prominent role. 

In future work, it would be interesting to test \eqref{eq:feq_Up-homcitss} for CIT over a wide range of $\langle \alpha_p \rangle$ and $\varphi$ values to determine whether the parameters in the model for $T^*_{L,1}$ should depend on these quantities. More generally, the complete model developed in Part~I should be tested for inhomogeneous particle-laden flows wherein the spatial transport terms play an important role. For example, the particle-laden channel flow of \cite{capecelatro2016strongly} would be a challenging test case.  In particular, for channel flows the correlated and uncorrelated particle velocity components generate separate spatial fluxes for all statistics.  From the model developed in \cite{capecelatro2016}, it is known that, depending on the Stokes number, one or the other of these fluxes may be dominant. As a result, the wall-normal distribution of  $\langle \alpha_p \rangle$, as well as other statistics, is very sensitive to how the spatial fluxes are modelled. In any case, as shown in this work, it can be expected that by including a stochastic model for $\mathbf{U}_s$ the resulting models for the spatial fluxes  will provide more robust closures for inhomogeneous turbulent particle-laden flows.

\section*{Acknowledgements}

R.O.F.\ gratefully acknowledges support from the U.S.\ National Science Foundation under Grants CBET-1437865 and ACI-1440443.

\appendix
\section{Fluid--particle limit}

The limit behaviour of the equations is only shown in homogeneous isotropic conditions for the sake of simplicity. In the limit case of tracer particles, i.e., $\tau_p \to 0$, we know from the equations for the stochastic model that $U_p \to U_s$ and $U_s \to U_f$, but we do not know if the model equation for $U_s$ is exactly the same as $U_p$. At the same time we have that the particle-phase uncorrelated velocity goes to zero, which is consistent. 
When the particle inertia becomes very small where $\beta_p = \beta_f \to 1$ and $k_{fp} = k_{f@p} \to k_f $, the particle-phase dissipation tends to $\varepsilon_p \to \varepsilon_f$, as we can see from \eqref{eq:epsilon_f} and \eqref{eq:epsilon_p}:
\begin{align}
\frac{d \varepsilon_f }{d t}  & = (C_{\epsilon 1} \mathcal{P} - C_{\epsilon 2} \varepsilon_f) \frac{\varepsilon_f}{k_f} +  C_{3} \frac{\varphi}{ \tau_p} \Bigl ( \frac{ k_{fp}}{k_{f@p}} \varepsilon_p - \beta_f \,  \varepsilon_f \Bigl ) ,
 \label{eq:epsilon_f}\\
\frac{d \varepsilon_p }{d t}  & = (C_{\epsilon 1} \mathcal{P} - C_{\epsilon 2} \varepsilon_p) \frac{\varepsilon_p}{k_p} +  C_{3} \frac{1}{ \tau_p} \Bigl ( \frac{ k_{fp}}{k_{f@p}} \varepsilon_f - \beta_p \, \varepsilon_p \Bigl ) .
 \label{eq:epsilon_p} 
\end{align}

Now we can check what happens to the stochastic equation for $U_s$. From the spatially homogeneous Lagrangian model, we can obtain
\begin{equation}
d \, U_s + \varphi d\,  U_p  = - \frac{1}{T_L}{ U}_{s}   \,dt  - \frac{\varphi}{T_{Lp}}{ U}_{p}   \,dt
+  \sqrt{C_0 \varepsilon_f}  \,  \, d{ W}_{s} +  \varphi \sqrt{C_{0p} \varepsilon_p}  \,  \, d{ W}_{p} .
\end{equation}
Now, when $\tau_p \to 0$, we can use one of the two equations to prove $d\, U_s = d\, U_p$, while the other two will give ${\delta} { v}_{p} \to 0$ and
\begin{equation}
d \, U_s = -  \frac{1}{T_L}  U_s \, dt + \frac{1}{(1 + \varphi)} \sqrt{C_0 \varepsilon_f}  \,  \, ( d{ W}_{s} + \varphi dW_p) .  \label{eq:limit-case}
\end{equation}
To obtain exactly the same equation as for one-way coupling, the white noise of the particle equation $dW_p$ should be replaced, in this limit, by the one employed in the fluid velocity equation $dW_s$.\footnote{In terms of the distribution function (i.e., weak convergence), the sum of two Wiener processes multiplied by constants is equivalent to a third Wiener process multiplied by the sum of the constants \citep{Klo_92}.} If this is not done, when the transport equation of the second-order moments is evaluated, i.e., $d \langle U_s^2 \rangle$, there will be a spurious term $- 2\varphi / (\varphi +1)^2 C_0 \varepsilon_f$ due to the fact that the two noises are uncorrelated. In any case, this term goes consistently to zero when the mass fraction $\varphi$ vanishes.

\bibliographystyle{jfm}
\bibliography{paper}

\begin{thebibliography}{33}
\expandafter\ifx\csname natexlab\endcsname\relax\def\natexlab#1{#1}\fi
\def\au#1{#1} \def\ed#1{#1} \def\yr#1{#1}\def\at#1{#1}\def\jt#1{\textit{#1}}
  \def\bt#1{#1}\def\bvol#1{\textbf{#1}} \def\vol#1{#1} \def\pg#1{#1}
  \def\publ#1{#1}\def\arxiv#1{#1}\def\org#1{#1}\def\st#1{\textit{#1}}

\bibitem[Ahmed \& Elghobashi(2000)]{ahmed2000mechanisms}
{\sc \au{Ahmed, A.~M.} \& \au{Elghobashi, S.}} \yr{2000}  \at{On the mechanisms
  of modifying the structure of turbulent homogeneous shear flows by dispersed
  particles}.  \jt{Phys. Fluids}  \bvol{12}~(11),  \pg{2906--2930}.

\bibitem[Balachandar \& Eaton(2010)]{balachandar2010turbulent}
{\sc \au{Balachandar, S.} \& \au{Eaton, J.~K.}} \yr{2010}  \at{Turbulent
  dispersed multiphase flow}.  \jt{Annu. Rev. Fluid Mech.}  \bvol{42},
  \pg{111--133}.

\bibitem[Balkovsky {\em et~al.\/}(2001)Balkovsky, Falkovich \&
  Fouxon]{balkovsky2001intermittent}
{\sc \au{Balkovsky, E.}, \au{Falkovich, G.} \& \au{Fouxon, A.}} \yr{2001}
  \at{Intermittent distribution of inertial particles in turbulent flows}.
  \jt{Phys. Rev. Lett.}  \bvol{86}~(13),  \pg{2790}.

\bibitem[Bosse {\em et~al.\/}(2006)Bosse, Kleiser \& Meiburg]{bosse2006small}
{\sc \au{Bosse, T.}, \au{Kleiser, L.} \& \au{Meiburg, E.}} \yr{2006}  \at{Small
  particles in homogeneous turbulence: Settling velocity enhancement by two-way
  coupling}.  \jt{Phys. Fluids}  \bvol{18}~(2),  \pg{027102}.

\bibitem[Capecelatro {\em et~al.\/}(2014)Capecelatro, Desjardins \&
  Fox]{capecelatro2014numerical}
{\sc \au{Capecelatro, J.}, \au{Desjardins, O.} \& \au{Fox, R.~O.}} \yr{2014}
  \at{Numerical study of collisional particle dynamics in cluster-induced
  turbulence}.  \jt{J. Fluid Mech.}  \bvol{747},  \pg{R2}.

\bibitem[Capecelatro {\em et~al.\/}(2015)Capecelatro, Desjardins \&
  Fox]{Fox2015}
{\sc \au{Capecelatro, J.}, \au{Desjardins, O.} \& \au{Fox, R.~O.}} \yr{2015}
  \at{On fluid-particle dynamics in fully developed cluster-induced
  turbulence.}  \jt{J. Fluid Mech.}  \bvol{780},  \pg{578--635}.

\bibitem[Capecelatro {\em et~al.\/}(2016{\natexlab{{\em a\/}}})Capecelatro,
  Desjardins \& Fox]{capecelatro2016strongly}
{\sc \au{Capecelatro, J.}, \au{Desjardins, O.} \& \au{Fox, R.~O.}}
  \yr{2016{\natexlab{{\em a\/}}}}  \at{Strongly coupled fluid-particle flows in
  vertical channels. i. reynolds-averaged two-phase turbulence statistics}.
  \jt{Phys. Fluids}  \bvol{28}~(3),  \pg{033306}.

\bibitem[Capecelatro {\em et~al.\/}(2016{\natexlab{{\em b\/}}})Capecelatro,
  Desjardins \& Fox]{capecelatro2016}
{\sc \au{Capecelatro, J.}, \au{Desjardins, O.} \& \au{Fox, R.~O}}
  \yr{2016{\natexlab{{\em b\/}}}}  \at{Strongly coupled fluid-particle flows in
  vertical channels. ii. turbulence modeling}.  \jt{Phys. Fluids}
  \bvol{28}~(3),  \pg{033307}.

\bibitem[Crowe {\em et~al.\/}(2011)Crowe, Schwarzkopf, Sommerfeld \&
  Tsuji]{crowe2011multiphase}
{\sc \au{Crowe, C.~T.}, \au{Schwarzkopf, J.~D.}, \au{Sommerfeld, M.} \&
  \au{Tsuji, Y.}} \yr{2011} {\em Multiphase flows with droplets and
  particles\/}.  \publ{CRC press}.

\bibitem[Dasgupta {\em et~al.\/}(1994)Dasgupta, Jackson \&
  Sundaresan]{dasgupta1994turbulent}
{\sc \au{Dasgupta, S.}, \au{Jackson, R.} \& \au{Sundaresan, S.}} \yr{1994}
  \at{Turbulent gas-particle flow in vertical risers}.  \jt{AIChE J.}
  \bvol{40}~(2),  \pg{215--228}.

\bibitem[Eaton \& Fessler(1994)]{eaton1994preferential}
{\sc \au{Eaton, J.~K.} \& \au{Fessler, J.~R.}} \yr{1994}  \at{Preferential
  concentration of particles by turbulence}.  \jt{Int. J. Multiphase Flow}
  \bvol{20},  \pg{169--209}.

\bibitem[Elghobashi \& Truesdell(1992)]{elghobashi1992direct}
{\sc \au{Elghobashi, S.} \& \au{Truesdell, G.~C.}} \yr{1992}  \at{Direct
  simulation of particle dispersion in a decaying isotropic turbulence}.
  \jt{J. Fluid Mech.}  \bvol{242},  \pg{655--700}.

\bibitem[Elghobashi \& Abou-Arab(1983)]{elghobashi1983two}
{\sc \au{Elghobashi, S.~E.} \& \au{Abou-Arab, T.~W.}} \yr{1983}  \at{A
  two-equation turbulence model for two-phase flows}.  \jt{The Physics of
  Fluids}  \bvol{26}~(4),  \pg{931--938}.

\bibitem[Ferrante \& Elghobashi(2003)]{ferrante2003physical}
{\sc \au{Ferrante, A.} \& \au{Elghobashi, S.}} \yr{2003}  \at{On the physical
  mechanisms of two-way coupling in particle-laden isotropic turbulence}.
  \jt{Phys. Fluids}  \bvol{15}~(2),  \pg{315--329}.

\bibitem[F\'evrier {\em et~al.\/}(2005)F\'evrier, Simonin \&
  Squires]{fevrier2005partitioning}
{\sc \au{F\'evrier, P.}, \au{Simonin, O.} \& \au{Squires, K.~D.}} \yr{2005}
  \at{Partitioning of particle velocities in gas--solid turbulent flows into a
  continuous field and a spatially uncorrelated random distribution:
  theoretical formalism and numerical study}.  \jt{J. Fluid Mech.}  \bvol{533},
   \pg{1--46}.

\bibitem[Forterre \& Pouliquen(2008)]{forterre2008flows}
{\sc \au{Forterre, Y.} \& \au{Pouliquen, O.}} \yr{2008}  \at{Flows of dense
  granular media}.  \jt{Annu. Rev. Fluid Mech.}  \bvol{40},  \pg{1--24}.

\bibitem[Fox(2014)]{Fox2014}
{\sc \au{Fox, R.~O.}} \yr{2014}  \at{On multiphase turbulence models for
  collisional fluid-particle flows.}  \jt{J. Fluid Mech.}  \bvol{742},
  \pg{368--424}.

\bibitem[Glasser {\em et~al.\/}(1998)Glasser, Sundaresan \&
  Kevrekidis]{glasser1998bubbles}
{\sc \au{Glasser, B.~J.}, \au{Sundaresan, S.} \& \au{Kevrekidis, I.~G.}}
  \yr{1998}  \at{From bubbles to clusters in fluidized beds}.  \jt{Phys. Rev.
  Lett.}  \bvol{81}~(9),  \pg{1849}.

\bibitem[Guazzelli \& Morris(2011)]{guazzelli2011physical}
{\sc \au{Guazzelli, E.} \& \au{Morris, J.~F.}} \yr{2011} {\em A physical
  introduction to suspension dynamics\/}, ,  \vol{vol.~45}.  \publ{Cambridge
  University Press}.

\bibitem[Hinze(1975)]{Hin_75}
{\sc \au{Hinze, J.O.}} \yr{1975} {\em Turbulence\/}, $2^{nd}$ edn.
  \publ{McGraw Hill, New-York}.

\bibitem[Kloeden \& Platen(1992)]{Klo_92}
{\sc \au{Kloeden, P.E.} \& \au{Platen, E.}} \yr{1992} {\em Numerical solution
  of stochastic differential equations\/}.  \publ{Springer-Verlag, Berlin}.

\bibitem[Minier {\em et~al.\/}(2014)Minier, Chibbaro \&
  B.]{minier2014guidelines}
{\sc \au{Minier, J.-P.}, \au{Chibbaro, S.} \& \au{B., Pope~S.}} \yr{2014}
  \at{Guidelines for the formulation of lagrangian stochastic models for
  particle simulations of single-phase and dispersed two-phase turbulent
  flows}.  \jt{Phys. Fluids}  \bvol{26}~(11),  \pg{113303}.

\bibitem[Minier {\em et~al.\/}(2004)Minier, Peirano \& Chibbaro]{Min_04}
{\sc \au{Minier, J-P.}, \au{Peirano, E.} \& \au{Chibbaro, S.}} \yr{2004}
  \at{Pdf model based on langevin equation for polydispersed two-phase flows
  applied to a bluff body-body gas-solid flow}.  \jt{Phys. Fluids}
  \bvol{16}~(7),  \pg{2419}.

\bibitem[Peirano {\em et~al.\/}(2006)Peirano, Chibbaro, Pozorski \&
  Minier]{Pei_06}
{\sc \au{Peirano, E.}, \au{Chibbaro, S.}, \au{Pozorski, J.} \& \au{Minier,
  J.-P.}} \yr{2006}  \at{Mean-field/pdf numerical approach for polydispersed
  turbulent two-phase flows}.  \jt{Prog. En. Comb. Sci.}  \bvol{32}~(3),
  \pg{315}.

\bibitem[Peirano \& Minier(2002)]{Pei_02}
{\sc \au{Peirano, E.} \& \au{Minier, J.-P.}} \yr{2002}  \at{Probabilistic
  formalism and hierarchy of models for polydispersed turbulent two-phase
  flows}.  \jt{Phys. Rev. E}  \bvol{65},  \pg{046301}.

\bibitem[Pope(1994)]{Pop_94a}
{\sc \au{Pope, S.~B.}} \yr{1994}  \at{On the relationship between stochastic
  lagrangian models of turbulence and second-order closures}.  \jt{Phys.
  Fluids}  \bvol{6}~(2),  \pg{973--985}.

\bibitem[Pope(2000)]{Pope_turbulent}
{\sc \au{Pope, S.~B.}} \yr{2000} {\em Turbulent flows\/}.  \publ{Cambridge
  University Press}.

\bibitem[Pozorski \& Minier(1998)]{Poz_98}
{\sc \au{Pozorski, J.} \& \au{Minier, J-P.}} \yr{1998}  \at{On the lagrangian
  turbulent dispersion models based on the langevin equation}.  \jt{Int. J.
  Multiphase Flow}  \bvol{24},  \pg{913--945}.

\bibitem[Puglisi(2014)]{puglisi2014transport}
{\sc \au{Puglisi, A.}} \yr{2014} {\em Transport and Fluctuations in Granular
  Fluids: From Boltzmann Equation to Hydrodynamics, Diffusion and Motor
  Effects\/}.  \publ{Springer}.

\bibitem[Stickel \& Powell(2005)]{stickel2005fluid}
{\sc \au{Stickel, J.~J} \& \au{Powell, R.~L.}} \yr{2005}  \at{Fluid mechanics
  and rheology of dense suspensions}.  \jt{Annu. Rev. Fluid Mech.}  \bvol{37},
  \pg{129--149}.

\bibitem[Sundaram \& Collins(1999)]{sundaram1999numerical}
{\sc \au{Sundaram, S.} \& \au{Collins, L.~R.}} \yr{1999}  \at{A numerical study
  of the modulation of isotropic turbulence by suspended particles}.  \jt{J.
  Fluid Mech.}  \bvol{379},  \pg{105--143}.

\bibitem[Tchen(1947)]{tchen1947mean}
{\sc \au{Tchen, C.M.}} \yr{1947}  \at{Mean value and correlation functions
  connected with the motion of small particles suspended in a turbulent fluid}.
   \jt{PhD Thesis, Delft} .

\bibitem[Viollet \& Simonin(1994)]{viollet1994modelling}
{\sc \au{Viollet, P.~L.} \& \au{Simonin, O.}} \yr{1994}  \at{Modelling
  dispersed two-phase flows: closure, validation and software development}.
  \jt{Appl. Mech. Rev}  \bvol{47}~(6),  \pg{S80--S84}.

\end{thebibliography}

\end{document}